\documentclass[11pt]{article}
\usepackage[utf8]{inputenc}
\usepackage[T1]{fontenc}
\usepackage{url}
\usepackage{cite}
\usepackage{amsmath}
\usepackage{amssymb}
\usepackage{amsthm}
\usepackage{algorithm}
\usepackage{algpseudocode}
\usepackage[colorlinks=true,linkcolor=blue,citecolor=blue,urlcolor=blue]{hyperref}

\newcommand{\IEEEPARstart}[2]{#1#2}
\newenvironment{IEEEkeywords}{\medskip\noindent\textbf{Keywords---}\space}{\medskip}
\providecommand{\markboth}[2]{}

\theoremstyle{plain}
\newtheorem{theorem}{Theorem}
\newtheorem{lemma}[theorem]{Lemma}
\newtheorem{proposition}[theorem]{Proposition}
\newtheorem{corollary}[theorem]{Corollary}
\newcommand{\restatedname}{}
\newtheorem*{restatedthm}{\restatedname}
\newenvironment{restated}[1]{\renewcommand{\restatedname}{Theorem~\ref{#1}}\begin{restatedthm}}{\end{restatedthm}}
\theoremstyle{definition}

\theoremstyle{remark}
\newtheorem{remark}[theorem]{Remark}

\newcommand{\F}{\mathbb{F}}
\newcommand{\GL}{\mathrm{GL}}
\newcommand{\OO}{\mathrm{O}}
\newcommand{\Sp}{\mathrm{Sp}}
\newcommand{\Ps}{\mathrm{Ps}}
\newcommand{\U}{\mathrm{U}}
\newcommand{\Stab}{\mathrm{Stab}}
\newcommand{\Hull}{\mathrm{Hull}}
\newcommand{\Rad}{\mathrm{Rad}}
\newcommand{\rank}{\mathrm{rank}}
\newcommand{\Cset}{\mathcal{C}}
\newcommand{\disc}{\mathrm{disc}}
\newcommand{\rowsp}{\mathrm{rowsp}}

\newcommand{\Nm}{\mathrm{N}}
\newcommand{\Gram}{\mathrm{Gram}}

\title{Uniform Sampling of Linear Codes with Prescribed Hull Dimension}

\author{Keita~Ishizuka\thanks{K.~Ishizuka is with the Information Technology R\&D Center, Mitsubishi Electric Corporation, Kamakura, Japan (e-mail: keitaishizuka1994@gmail.com).}\thanks{This work was supported by JSPS KAKENHI Grant Number JP25K17290.}}

\begin{document}
\maketitle

\begin{abstract}
We give exactly uniform samplers for linear codes with prescribed hull dimension, for the Euclidean, Hermitian, and symplectic inner products and over finite fields of every characteristic.
Except for the Euclidean form over fields of odd characteristic, these are the first uniform samplers for arbitrary prescribed hull dimension.
Each sample costs an expected \(O(k^2 N + k \log q)\) field operations for codes of dimension \(k\) in a space of dimension \(N\) over \(\mathbb{F}_q\) or \(\mathbb{F}_{q^2}\).
\end{abstract}

\begin{IEEEkeywords}
Linear codes, hull dimension, uniform sampling, classical groups, Witt's theorem, LCD codes, self-orthogonal codes, entanglement-assisted quantum codes, code-based cryptography.
\end{IEEEkeywords}

\section{Introduction}\label{sec:intro}

\IEEEPARstart{U}{p} to a scalar, a non-degenerate reflexive form on a finite-dimensional vector space is symmetric, alternating, or Hermitian~\cite[Thm.~7.1]{Taylor}.
On codes these give the Euclidean, symplectic, and Hermitian inner products, and for each of them the \emph{hull} $\Hull(C) := C \cap C^\perp$ of a code $C$ plays an important role in coding theory, cryptography, and quantum error correction.
Binary Euclidean LCD codes ($\Hull(C) = 0$) are used in countermeasures against side-channel and fault-injection attacks~\cite{CarletGuilley2016}.
Euclidean self-orthogonal codes ($\Hull(C) = C$) are needed for instance generation in the code-based signature scheme SPECK~\cite{SPECK}, and codes with a prescribed Euclidean hull dimension are the central primitive of the Hollow~LWE encryption scheme~\cite{HollowLWE}.
Codes in $\F_q^{2n}$ with symplectic hull dimension $\ell$ are the check spaces of entanglement-assisted (EA) stabilizer codes on $n$ qudits that consume $(\dim C - \ell)/2$ ebits, and codes in $\F_{q^2}^n$ with Hermitian hull dimension $\ell$ give EA stabilizer codes consuming $\dim C - \ell$ ebits~\cite{BDH06,WB08} for qubits and~\cite{GHMR19} for all $q$; the self-orthogonal case is that of ordinary stabilizer codes~\cite{CRSS98,AK01,KKKS06}.
Uniform sampling of such codes is useful in each of these settings: it offers a complementary approach to the classification of codes, it drives Monte Carlo simulation of quantum codes, and in Hollow~LWE and SPECK the uniform sampling of codes with prescribed Euclidean hull dimension is part of the scheme itself.

For the Euclidean inner product, Albrecht, Ben\v{c}ina, and Lai gave a partial result in Hollow~LWE~\cite{HollowLWE}: their sampler grows the hull one dimension at a time and is proved uniform when $2\ell \le k$, where $k$ is the dimension and $\ell$ the hull dimension.
The same authors later gave a sampler for odd $q$ that, for $n$ and $q$ polynomial in $k$, aborts with negligible probability and is otherwise exactly uniform for a prescribed hull dimension and square class of the discriminant of $C/\Hull(C)$~\cite[Thm.~28]{ABL26}, with the mass formula of Li, Shi, and Ling~\cite{LiShiLing2025} used in the analysis; for a prescribed hull dimension alone, it picks the square class uniformly at random and is within total variation distance $q^{-(k-\ell)/2}$ of uniform~\cite[Cor.~30]{ABL26}.
The case of even $q$ was left open.
For self-orthogonal codes and odd $q$, an $\varepsilon$-uniform sampler was given in the conference version of the present paper~\cite{ITWversion}.
For the Hermitian and symplectic inner products, uniform samplers are known only for special hull dimensions.
Uniformly random symplectic self-orthogonal codes, that is, stabilizer codes, are obtained by applying a uniformly random Clifford operator to a fixed stabilizer code, and the $n$-qubit Clifford group can be sampled exactly~\cite{KS14,BM21}.
No uniform sampler was available for a prescribed Hermitian or symplectic hull dimension, that is, for EA codes with a prescribed number of ebits.
Table~\ref{tab:prior} summarizes the situation.

\begin{table}[h]
\centering
\caption{Exactly uniform samplers of codes with prescribed hull dimension $\ell$ and dimension $k$.}\label{tab:prior}
\medskip
\begin{tabular}{lll}
\hline
inner product & previous work & this work \\
\hline
symplectic & $\ell = k$, $q = 2$~\cite{KS14,BM21} & all $\ell$, all $q$ \\
Hermitian & none & all $\ell$, all $q$ \\
Euclidean, $q$ odd & all $\ell$, per square class~\cite{ABL26} & all $\ell$ \\
Euclidean, $q$ even & none & all $\ell$ \\
\hline
\end{tabular}
\end{table}

\subsection{Our contribution}

We give an exactly uniform sampler of codes with prescribed hull dimension for the Euclidean inner product (both for odd and for even $q$), the Hermitian inner product, and the symplectic inner product.
Except for the Euclidean inner product with $q$ odd, which was treated in~\cite{ABL26} up to the weighting of the square classes, these are the first exactly uniform samplers for arbitrary prescribed hull dimension.

The algorithm is simple and general: a single framework covers all three inner products and every $q$.
For each inner product, the isometry group $G$ (the orthogonal, pseudo-symplectic, unitary, or symplectic group) acts on the codes of given length, dimension, and hull dimension with at most four orbits.
Once a simple representative of each orbit is fixed, the problem therefore reduces to the uniform sampling of an element of $G$, and any sampler of $G$ can be used for this step.
By making the subgroup algorithm of Diaconis and Shahshahani~\cite{DS87} explicit for our groups and accelerating it, we obtain exactly uniform samplers that take an expected $O(k^2 N + k \log q)$ operations, where $k$ is the dimension of the code and $N$ the dimension of the ambient space ($N = n$ for the Euclidean and Hermitian forms and $N = 2n$ for the symplectic form).
We also provide an implementation of all samplers in Python~\cite{hullsampler}.

Compared with previous work, the samplers have three advantages.
They cover the Hermitian and symplectic forms and the Euclidean form with $q$ even, which were left open.
For the Euclidean form with $q$ odd, their expected cost is proved for all $n$, $k$, and $q$, whereas the bounds on the abort probability and running time of the sampler of~\cite{ABL26} assume that $n$ and $q$ are polynomial in $k$.
They are faster: the sampler of~\cite{ABL26} is proved to run in time $O(k(n^{\omega} + \log^2 q))$ for its LCD part and $O(n^{\omega} + n k q \log^4 q + k^2)$ for each hull vector, where $\omega$ is the exponent of matrix multiplication~\cite[Lems.~17, 24]{ABL26}, whereas ours costs an expected $O(k^2 n + k \log q)$ operations in total (Corollary~\ref{cor:euclid-sampler}).

The rest of the paper is organized as follows.
Section~\ref{sec:prelim} recalls the three inner products, Witt's theorem, the Anzahl theorems and their relation to the mass formulas of coding theory together with the two counts we use, facts on finite fields, and the facts on uniform sampling that we use.
Section~\ref{sec:framework} presents the framework: Algorithm~\ref{alg:main} reduces the uniform sampling of codes to (i) the orbits of the isometry group and their sizes, (ii) a representative of each orbit with a simple generator matrix, and (iii) a uniform sampler of the isometry group.
Sections~\ref{sec:symp}--\ref{sec:even} provide these items for the symplectic form (one orbit), the Hermitian form (one orbit), the Euclidean form with $q$ odd (at most two orbits), and the Euclidean form with $q$ even (at most four orbits); we begin with the symplectic form, since it is the easiest case and the case of even $q$ builds on it.
Section~\ref{sec:applications} discusses applications and Section~\ref{sec:conclusion} concludes.

\section{Preliminaries}\label{sec:prelim}

\subsection{Three inner products and their hulls}\label{ssec:codes}

We consider three spaces $(V,B)$, each with a non-degenerate form $B$:
\begin{itemize}
\item \emph{Euclidean:} $V = \F_q^n$ and $B(x,y) := \sum_{i=1}^n x_i y_i$, symmetric and bilinear.
\item \emph{Hermitian:} $V = \F_{q^2}^n$ and $B(x,y) := \sum_{i=1}^n x_i \bar y_i$ with $\bar y := y^q$, sesquilinear with $B(y,x) = \overline{B(x,y)}$.
\item \emph{Symplectic:} $V = \F_q^{2n}$ and $B((a|b),(a'|b')) := a \cdot b' - b \cdot a'$ for $a,b,a',b' \in \F_q^n$, bilinear and alternating, $B(x,x) = 0$.
\end{itemize}
A \emph{(linear) code} is a subspace $C \le V$: codes are $\F_q$-linear in the Euclidean and symplectic cases and $\F_{q^2}$-linear in the Hermitian case; we write $\F$ for the field of scalars and $N := \dim_\F V$, so $N = n$, $n$, $2n$ respectively.
We write $b_1, \ldots, b_N$ for the standard basis of $V$; in the symplectic case $b_i$ and $b_{n+i}$ $(i \le n)$ are the $i$-th unit vectors of the $a$-part and of the $b$-part of $(a|b)$, so that $B(b_i, b_{n+j}) = \delta_{ij}$ and $B(b_i,b_j) = B(b_{n+i},b_{n+j}) = 0$.
A code of dimension $k$ is given by a generator matrix $G \in \F^{k \times N}$ with $C = \rowsp(G)$.

In all three cases $B(x,y) = 0$ if and only if $B(y,x) = 0$, so the \emph{dual code} $C^\perp := \{v \in V : B(v,c) = 0 \ \forall c \in C\}$ is well defined, of dimension $N - \dim C$, and the \emph{hull} is $\Hull(C) := C \cap C^\perp$.
The code $C$ is \emph{LCD} if $\Hull(C) = 0$ and \emph{self-orthogonal} if $\Hull(C) = C$.

\begin{remark}\label{rem:terms}
The hull is the radical of the restricted form, $\Hull(C) = \Rad(C) := \{c \in C : B(c, C) = 0\}$, and $C$ is LCD or self-orthogonal exactly when $B|_C$ is non-degenerate or zero. From here on we use the terms of the theory of forms (radical, non-degenerate, totally isotropic) when we work with $B|_C$, and keep the terms of coding theory for statements about codes.
\end{remark}

Our object is the family
\[
\Cset_{n,k,\ell} \;:=\; \{ C \le V : \dim C = k,\; \dim \Hull(C) = \ell \},
\]
written $\Cset^{\mathrm{E}}_{n,k,\ell}$, $\Cset^{\mathrm{H}}_{n,k,\ell}$, $\Cset^{\mathrm{S}}_{n,k,\ell}$ when the inner product has to be named.
The Euclidean family contains the LCD and self-orthogonal codes used in side-channel countermeasures and code-based cryptography, and the Hermitian and symplectic families are the check spaces of EA stabilizer codes with $k - \ell$ and $(k-\ell)/2$ ebits (Section~\ref{sec:applications}).
The \emph{Gram matrix} of vectors $c_1, \ldots, c_k$ is $\Gram(c_1, \ldots, c_k) := (B(c_i,c_j))_{i,j}$.

\begin{lemma}[Hull dimension and admissibility]\label{lem:basic}
Let $C \le V$ have basis $c_1, \ldots, c_k$ and hull dimension $\ell$.
\begin{enumerate}
\item $\ell = k - \rank \Gram(c_1, \ldots, c_k)$.
\item $B$ induces a non-degenerate form $\bar B_C(c + \Rad(C), c' + \Rad(C)) := B(c,c')$ on the quotient $\bar C := C/\Rad(C)$, of the same kind as $B$; in particular $k - \ell$ is even in the symplectic case.
\item $k + \ell \le N$.
\end{enumerate}
\end{lemma}
\begin{proof}
(1) The vector $\sum_i a_i c_i$ lies in $C^\perp$ exactly when $\sum_i a_i B(c_i, c_j) = 0$ for all $j$, so $\Rad(C) = \Hull(C)$ corresponds to the left kernel of the Gram matrix.
(2) For $h \in \Rad(C)$ and $c \in C$ we have $B(h,c) = B(c,h) = 0$, so $\bar B_C$ is well defined, and $\bar B_C(c + \Rad(C), \cdot) \equiv 0$ forces $c \in \Rad(C)$; a non-degenerate alternating form exists only in even dimension.
(3) $\dim(C + C^\perp) = k + (N - k) - \ell = N - \ell$, and $C \subseteq C + C^\perp$.
\end{proof}

We call $(\bar C, \bar B_C)$ the \emph{non-degenerate quotient} of $C$; it has dimension $k - \ell$.
For every complement $W$ of $\Rad(C)$ in $C$, $C = \Rad(C) \oplus W$, the space $(W, B|_W)$ is non-degenerate and isometric to $(\bar C, \bar B_C)$ via $w \mapsto w + \Rad(C)$, by Lemma~\ref{lem:basic}(2).

\subsection{Isometry groups and Witt's theorem}\label{ssec:witt}

An injective linear map $\phi: C \to C'$ between subspaces is an \emph{isometry} if $B(\phi x, \phi y) = B(x,y)$ for all $x, y \in C$.
The isometry group of $(V,B)$ is $G := \{ g \in \GL(V) : B(xg,yg) = B(x,y) \ \forall x,y \}$: the orthogonal group $\OO(n,q)$, the unitary group $\U(n,q^2)$, and the symplectic group $\Sp(2n,q)$ in the three cases; for $q$ even the Euclidean case is special and is treated in Section~\ref{sec:even}.
We let $g$ act on row vectors from the right and write $Cg := \{cg : c \in C\}$, so that $\rowsp(G_0)g = \rowsp(G_0 g)$.
Since $B(xg,yg) = B(x,y)$, we have $(Cg)^\perp = C^\perp g$ and $\Hull(Cg) = \Hull(C)g$, so $G$ acts on $\Cset_{n,k,\ell}$.

\begin{theorem}[Witt's extension theorem~{\cite[Thm.~7.4]{Taylor}}]\label{thm:witt}
Let $(V,B)$ be a non-degenerate space whose form is alternating, Hermitian, or symmetric over a field of odd characteristic.
Every isometry between two subspaces of $V$ extends to an element of the isometry group of $(V,B)$.
\end{theorem}

Theorem~\ref{thm:witt} fails for symmetric forms in characteristic $2$ that are not alternating, which is why the Euclidean form with $q$ even is treated separately.
Together with the well-known fact that a degenerate form is the orthogonal sum of its radical and a non-degenerate form~\cite[Ch.~7]{Taylor}, it reduces transitivity on $\Cset_{n,k,\ell}$ to the classification of non-degenerate forms of dimension $k - \ell$; we record this standard consequence in the form we use.

\begin{lemma}[Transitivity criterion]\label{lem:transitive}
In the setting of Theorem~\ref{thm:witt}, let $C, C' \in \Cset_{n,k,\ell}$.
If the non-degenerate quotients $(\bar C, \bar B_C)$ and $(\bar C', \bar B_{C'})$ are isometric, then $Cg = C'$ for some $g \in G$.
\end{lemma}
\begin{proof}
Choose complements $W$ of $\Rad(C)$ in $C$ and $W'$ of $\Rad(C')$ in $C'$; they are isometric to the non-degenerate quotients, hence to each other.
Let $\phi_W: W \to W'$ be an isometry and $\phi_H: \Rad(C) \to \Rad(C')$ any linear bijection.
Then $\phi := \phi_H \oplus \phi_W: C \to C'$ is an isometry: $B$ vanishes on $\Rad(C) \times C$ and on $\Rad(C') \times C'$, so only the $W$-block contributes.
Theorem~\ref{thm:witt} extends $\phi$ to some $g \in G$, and $Cg = C'$.
\end{proof}

For a totally isotropic subspace $H$ of dimension $\ell$, a \emph{Witt decomposition} is $V = H \oplus M \oplus H^*$, where $H^*$ is totally isotropic of dimension $\ell$ with bases $h_1, \ldots, h_\ell$ of $H$ and $h_1^*, \ldots, h_\ell^*$ of $H^*$ satisfying $B(h_i, h_j^*) = \delta_{ij}$, and $M := (H \oplus H^*)^\perp$; such $H^*$ exists by~\cite[Lemma~7.5]{Taylor}, and $M$ is non-degenerate and isometric to $H^\perp/H$ because $H^\perp = H \oplus M$.

\subsection{Anzahl theorems and mass formulas}\label{ssec:anzahl}

A code $C \in \Cset_{n,k,\ell}$ is a $k$-dimensional subspace whose restricted form has a radical of dimension $\ell$ and induces on $C/\Rad(C)$ the form $\bar B_C$ of Lemma~\ref{lem:basic}(2), so its isometry type is determined by $(k, \ell)$ and the isometry class of $\bar B_C$.
Counting the subspaces of a finite non-degenerate space by the isometry type of the restricted form is a classical problem of finite geometry~\cite[\S 3]{Wan97}.
Segre counted the totally singular subspaces of a quadric, and Pless counted the totally isotropic subspaces of a symplectic space.
In 1964--1966 Wan and his students Dai, Feng, and Yang determined the sizes of all orbits of subspaces under the symplectic, unitary, and orthogonal groups, and called these results the Anzahl theorems; Liu and Wan later did the same for the pseudo-symplectic group~\cite{LiuWan}.
The results are compiled in~\cite{Wan} and surveyed in~\cite{Wan97}.
For the inner products of this paper, they give the number of codes of each dimension, hull dimension, and isometry type.

In coding theory, the same numbers were later obtained independently, under the name of mass formulas and in different terminology and notation.
Sendrier~\cite{Sendrier97} expressed the number of codes with a given hull dimension through the numbers of self-orthogonal codes, Carlet et al.~\cite{CMTQ19} counted LCD codes for $q = 2$ and for odd $q$ through the action of the orthogonal group, and Li, Shi, and Ling~\cite{LiShiLing2025,LSLL24} counted codes with arbitrary prescribed hull dimension for the Euclidean, Hermitian, and symplectic forms.
In this paper, orbit sizes are needed only for the weights of Algorithm~\ref{alg:main} when there is more than one orbit.
We take them from the mass formulas of coding theory where these give the counts we need, since their notation is closer to ours; for $q$ even, the split of $\Cset_{n,k,\ell}$ into the four orbits of Section~\ref{sec:even} is not stated there for $\ell > 0$, and we use the Anzahl numbers.
The history above is recalled to acknowledge the Anzahl theorems.

The counts we use are stated where they are needed: the symplectic Anzahl theorem in Section~\ref{sec:symp}, the mass formula of Li, Shi, and Ling for $q$ odd in Section~\ref{sec:euclid-odd}, and the pseudo-symplectic Anzahl numbers in Section~\ref{sec:even}.

\subsection{Finite fields}\label{ssec:ff}

We collect the facts about finite fields used in Sections~\ref{sec:herm} and~\ref{sec:euclid-odd}.
The group $\F_q^\times$ is cyclic of order $q-1$~\cite[Ch.~7, \S 1, Thm.~1]{IrelandRosen}.
For $q$ odd, let $\eta$ be the quadratic character of $\F_q$, extended by $\eta(0) := 0$, and $(\F_q^*)^2$ the subgroup of non-zero squares.
Since $\F_q^\times$ is cyclic, for $q$ odd $(\F_q^*)^2$ is its unique subgroup of index $2$, and $c \in \F_q^*$ is a square if and only if $c^{(q-1)/2} = 1$ (Euler's criterion), so $\eta(c) = c^{(q-1)/2}$ is computed in $O(\log q)$ operations by repeated squaring.

\begin{lemma}\label{lem:minus-one-square}
For an odd prime power $q$, the element $-1$ is a square in $\F_q^*$ if and only if $q \equiv 1 \pmod 4$.
\end{lemma}
\begin{proof}
By Euler's criterion, $-1$ is a square if and only if $(-1)^{(q-1)/2} = 1$, that is, $4 \mid q-1$.
\end{proof}

\begin{lemma}\label{lem:sum-two-squares}
For any odd prime power $q$ and any $c \in \F_q$, there exist $a, b \in \F_q$ such that $a^2 + b^2 = c$.
\end{lemma}
\begin{proof}
The sets
\[
S_1 := \{a^2 : a \in \F_q\}, \qquad S_2 := \{c - b^2 : b \in \F_q\}
\]
each have size $(q+1)/2$.
Since $|S_1| + |S_2| = q + 1 > |\F_q|$, the two sets intersect by pigeonhole, yielding $a, b$ with $a^2 = c - b^2$, i.e., $a^2 + b^2 = c$.
\end{proof}

For the field $\F_{q^2}$ with conjugation $\bar a := a^q$, the norm $\Nm(a) := a \bar a = a^{q+1}$ maps $\F_{q^2}^*$ onto $\F_q^*$~\cite[Lemma~10.1(v)]{Taylor}, so each value is taken by exactly $q+1$ elements.
Moreover, $\xi \mapsto \xi^{q-1}$ maps $\F_{q^2}^*$ onto $\ker \Nm$, and every element of $\ker\Nm$ is the image of exactly $q-1$ elements: $\Nm(\xi^{q-1}) = \xi^{q^2-1} = 1$, and in the cyclic group $\F_{q^2}^*$ of order $q^2-1$ the image of the $(q-1)$-th power map has order $q+1 = |\ker\Nm|$.

\begin{lemma}[Cipolla's method~{\cite[\S 2]{DoliskaniSchost}}]\label{lem:cipolla}
Let $q$ be odd and $a \in (\F_q^*)^2$.
If $b \in \F_q$ is such that $b^2 - 4a$ is a non-square, then $f(Y) := Y^2 - bY + a$ is irreducible over $\F_q$, and $Y^{(q+1)/2} \bmod f$ is a square root of $a$.
Drawing $b$ uniformly from $\F_q$ gives such a $b$ after an expected $O(1)$ draws, so a square root costs an expected $O(\log q)$ operations in $\F_q$.
\end{lemma}

\subsection{Uniform sampling}\label{ssec:prob}

All sets in this paper are finite.
For a finite set $X$ we write $u_X$ for the uniform distribution on $X$, and for distributions $\mu, \nu$ on $X$ we use the total variation distance $d(\mu,\nu) := \max_{A \subseteq X} |\mu(A) - \nu(A)| = \frac12 \sum_{x} |\mu(x) - \nu(x)|$.
A randomized algorithm with output distribution $\mu$ on $X$ is an \emph{exactly uniform} sampler of $X$ if $\mu = u_X$, and an \emph{$\varepsilon$-uniform} sampler if $d(\mu, u_X) \le \varepsilon$.
For a map $f: X \to Y$, the \emph{push-forward} $f_*\mu$ is the distribution of $f(x)$ for $x \sim \mu$, that is, $f_*\mu(y) = \mu(f^{-1}(y))$.

\begin{lemma}[Push-forward and mixtures]\label{lem:pushforward}
Let $\mu, \nu$ be distributions on $X$.
\begin{enumerate}
\item For every map $f: X \to Y$, $d(f_*\mu, f_*\nu) \le d(\mu,\nu)$.
\item For distributions $\mu_o, \nu_o$ on $X$ and weights $p^o \ge 0$ with $\sum_o p^o = 1$, $d(\sum_o p^o \mu_o, \sum_o p^o \nu_o) \le \sum_o p^o d(\mu_o, \nu_o)$.
\item If a finite group $G$ acts transitively on $X$ and $x_0 \in X$, then the orbit map $\phi: G \to X$, $g \mapsto x_0 g$, satisfies $\phi_* u_G = u_X$; hence if $\mu$ is an $\varepsilon$-uniform distribution on $G$, then $\phi_*\mu$ is $\varepsilon$-uniform on $X$, and in particular $\phi_*\mu = u_X$ if $\mu = u_G$.
\item If $X = \bigsqcup_o X_o$ with non-empty $X_o$ and $p^o := |X_o|/|X|$, then $\sum_o p^o u_{X_o} = u_X$.
\end{enumerate}
\end{lemma}
\begin{proof}
(1) $f_*\mu(A) - f_*\nu(A) = \mu(f^{-1}(A)) - \nu(f^{-1}(A))$ for every $A \subseteq Y$.
(2) $\frac12 \sum_x |\sum_o p^o (\mu_o(x) - \nu_o(x))| \le \sum_o p^o \cdot \frac12 \sum_x |\mu_o(x) - \nu_o(x)|$.
(3) The elements $g$ with $x_0 g = x_0 h$ form the coset $\Stab_G(x_0)\, h$, so every element of $X$ is the image of exactly $|\Stab_G(x_0)|$ elements of $G$; and $d(\phi_*\mu, u_X) = d(\phi_*\mu, \phi_* u_G) \le d(\mu, u_G)$ by (1).
(4) For $x \in X_o$ the left-hand side takes the value $p^o / |X_o| = 1/|X|$.
\end{proof}

\section{The Framework}\label{sec:framework}

Let $\Cset_{n,k,\ell} = \bigsqcup_{o \in \mathcal{O}} \Cset^{o}_{n,k,\ell}$ be the decomposition into $G$-orbits, and suppose that we have (i) the orbit sizes, (ii) a generator matrix $G_0^{o}$ of a representative of each non-empty orbit, and (iii) a sampler of $G$.
Algorithm~\ref{alg:main} combines them.

\begin{algorithm}[t]
\caption{Uniform sampler for $\Cset_{n,k,\ell}$}\label{alg:main}
\begin{algorithmic}[1]
\Require an inner product, parameters $(n,k,\ell,q)$ with $\Cset_{n,k,\ell} \ne \emptyset$
\Ensure exactly uniform $C \in \Cset_{n,k,\ell}$
\State \textbf{Preprocessing (once):} for each non-empty orbit $o$, construct $G_0^{o}$ and $p^{o} := |\Cset^{o}_{n,k,\ell}| / |\Cset_{n,k,\ell}|$
\State draw an orbit $o$ with probability $p^{o}$
\State draw $g \in G$ from the sampler of $G$
\State \Return $C := \rowsp(G_0^{o} \, g)$
\end{algorithmic}
\end{algorithm}

\begin{theorem}[Uniformity]\label{thm:uniformity}
Given (i) and (ii), if the sampler of $G$ in (iii) is $\varepsilon$-uniform, then the output of Algorithm~\ref{alg:main} is $\varepsilon$-uniform on $\Cset_{n,k,\ell}$; in particular, an exactly uniform sampler of $G$ gives an exactly uniform sampler of $\Cset_{n,k,\ell}$.
Each call costs one draw of $g$ plus $O(kN^2)$ operations in $\F$.
\end{theorem}
\begin{proof}
Let $\mu$ be the distribution of $g$ and $\phi_o: G \to \Cset^{o}_{n,k,\ell}$, $g \mapsto \rowsp(G_0^{o} g)$, the orbit map of the representative of $o$.
By Lemma~\ref{lem:pushforward}(3), $(\phi_o)_* u_G$ is the uniform distribution on $\Cset^o_{n,k,\ell}$, so by Lemma~\ref{lem:pushforward}(4) and the choice of $p^o$ the uniform distribution on $\Cset_{n,k,\ell}$ is $\sum_o p^o (\phi_o)_* u_G$.
The output has distribution $\sum_o p^o (\phi_o)_*\mu$, so by Lemma~\ref{lem:pushforward}(2) and~(1),
\begin{align*}
d\Bigl(\sum_o p^o (\phi_o)_*\mu,\ \sum_o p^o (\phi_o)_* u_G\Bigr) &\le \sum_o p^o\, d\bigl((\phi_o)_*\mu, (\phi_o)_* u_G\bigr) \\
&\le \sum_o p^o\, d(\mu, u_G) \;=\; d(\mu, u_G) \;\le\; \varepsilon.
\end{align*}
The product $G_0^{o} g$ costs $O(kN^2)$.
\end{proof}

When there is a single orbit, item (i) is not needed and line~2 disappears.
In the rest of this paper we provide, for each inner product, (i) the orbits and their sizes, (ii) a standard representative of each orbit, and (iii) a uniform sampler of $G$; for (iii) we use the subgroup algorithm of Diaconis and Shahshahani, which we now recall.

\subsection{Uniform sampling of the isometry group}\label{ssec:group-sampling}

By Theorem~\ref{thm:uniformity}, any sampler of $G$ can be used in line~3, and uniform sampling of group elements can be done in several ways.
For a group given only by generators, black-box methods such as Babai's algorithm~\cite{Babai91} and the Cooperman--Dixon cube construction~\cite{Cooperman,Dixon2008} return approximately uniform elements, and the product replacement algorithm~\cite{CLMNO95} is a practical method for the same purpose.
For the classical groups over $\F_q$, the subgroup algorithm of Diaconis and Shahshahani~\cite{DS87}, which they work out for $\mathrm{GL}_n(\F_q)$ along the chain of stabilizers of the successive basis vectors~\cite[\S 2]{DS87}, gives exact samplers tailored to the group, which are used in Sections~\ref{sec:symp}--\ref{sec:even}.
We state the subgroup algorithm in the form we use; Witt's theorem is what makes each step uniform.

\begin{lemma}[Uniform isometry, one vector at a time; cf.~{\cite[\S 2]{DS87}}]\label{lem:seq}
In the setting of Theorem~\ref{thm:witt}, let $G$ be the isometry group of $(V,B)$ and $v_1, \ldots, v_s$ linearly independent vectors of $V$, $s \le N$.
Draw $w_1, \ldots, w_s$ successively, $w_i$ uniformly from
\begin{equation}\label{eq:seq}
\begin{split}
\Sigma_i \;:=\; \bigl\{\, x \in V \setminus \langle w_1, \ldots, w_{i-1}\rangle \;:\;{}& B(w_j, x) = B(v_j, v_i)\ (j < i), \\ & B(x,x) = B(v_i,v_i) \,\bigr\}.
\end{split}
\end{equation}
Then each $\Sigma_i$ is non-empty with $|\Sigma_i|$ depending only on $i$, and $(w_1, \ldots, w_s)$ is distributed as $(v_1 g, \ldots, v_s g)$ for a uniform $g \in G$.
In particular, for $s = N$ the map $g$ with $v_i g = w_i$ is an exactly uniform element of $G$.
\end{lemma}
\begin{proof}
Suppose that $v_j \mapsto w_j$ $(j < i)$ is an isometry, which holds for $i = 1$.
By Theorem~\ref{thm:witt} it extends to some $h \in G$, and $h$ maps the set~\eqref{eq:seq} computed for $(v_1, \ldots, v_{i-1})$ bijectively onto the set computed for $(w_1, \ldots, w_{i-1})$; the former contains $v_i$, so $\Sigma_i$ is non-empty and $|\Sigma_i|$ depends only on $i$.
Any $w_i \in \Sigma_i$ extends the isometry to $v_1, \ldots, v_i$.
Hence the procedure produces exactly the tuples $(v_1 g, \ldots, v_s g)$, $g \in G$, each with probability $\prod_i |\Sigma_i|^{-1}$.
The elements $g$ giving the same tuple $(v_1 g, \ldots, v_s g)$ form a coset of the pointwise stabilizer of $v_1, \ldots, v_s$, so the map $g \mapsto (v_1 g, \ldots, v_s g)$ pushes $u_G$ forward to the uniform distribution on these tuples, as in Lemma~\ref{lem:pushforward}(3).
\end{proof}

The only step of Lemma~\ref{lem:seq} that depends on the group is the uniform draw from $\Sigma_i$, which Sections~\ref{sec:symp}--\ref{sec:even} carry out explicitly.
Each $\Sigma_i$ is cut out by a linear system and at most one further condition on $B(x,x)$, which we meet by rejection and rescaling; the linear part is handled by the following lemma.

\begin{lemma}[Cost of the draws]\label{lem:cost}
Let $w_1, \ldots, w_{i-1} \in V$ and $c_1, \ldots, c_{i-1} \in \F$, and keep the system $B(w_j, x) = c_j$ $(j < i)$ in reduced row echelon form.
A uniform solution $x$ costs $O(iN)$ operations, and adding the equation of a further vector $w_i$ costs $O(iN)$.
Hence, if each $w_i$ is obtained from an expected $O(1)$ uniform solutions of the system for $j < i$ and $O(N)$ further operations, then $w_1, \ldots, w_s$ cost an expected $\sum_{i \le s} O(iN) = O(s^2 N)$ operations.
\end{lemma}
\begin{proof}
The map $x \mapsto B(w_j, x)$ is linear, and in the Hermitian case $x \mapsto \overline{B(w_j,x)} = \sum_t \bar w_{j,t} x_t$ is linear, so the conditions form a linear system of $i-1$ equations in $N$ unknowns.
In reduced row echelon form, a uniform solution is obtained by choosing the non-pivot coordinates uniformly and solving for each of the at most $i - 1$ pivot coordinates with $O(N)$ operations.
A new equation is reduced against the existing rows in $O(iN)$ operations and, if it has a new pivot, eliminated from them in $O(iN)$ operations.
\end{proof}

In particular, the whole of $g$ costs an expected $O(N^3)$ operations in all our cases.

\subsection{Truncation}

With the group samplers of Sections~\ref{sec:symp}--\ref{sec:even}, a call of Algorithm~\ref{alg:main} costs $O(N^3)$, dominated by line~3.
Line~4, however, uses only a few rows of $g$.

\begin{proposition}[Truncation]\label{prop:trunc}
Let $v_1, \ldots, v_N$ be a basis of $V$, and suppose that the sampler of $G$ in line~3 draws vectors $w_1, w_2, \ldots, w_N$ in this order and returns the map $g$ with $v_i g = w_i$, and that $g$ is exactly uniform on $G$.
Write $G_0^{o} = A V_s$ with $A \in \F^{k \times s}$ and $V_s$ the matrix with rows $v_1, \ldots, v_s$, and let $W_s$ be the matrix with rows $w_1, \ldots, w_s$.
Then $\rowsp(A W_s)$ is distributed as $\rowsp(G_0^{o} h)$ for a uniform $h \in G$; so the sampler can be stopped after $w_s$ without changing the output distribution of Algorithm~\ref{alg:main}.
\end{proposition}
\begin{proof}
The first $s$ draws are made before the others, so stopping after $w_s$ does not change their distribution, which is that of $(v_1 h, \ldots, v_s h)$ for a uniform $h \in G$.
The code $\rowsp(A W_s)$ is a function of $(w_1, \ldots, w_s)$, and it equals $\rowsp(A V_s h) = \rowsp(G_0^o h)$ when $(w_1, \ldots, w_s) = (v_1 h, \ldots, v_s h)$; push forward by Lemma~\ref{lem:pushforward}.
\end{proof}

In Sections~\ref{sec:symp}--\ref{sec:even} the representatives involve $s \le k + \ell + 4$ basis vectors, and the group samplers draw the images of the basis vectors in an order that starts with them.
Since the $i$-th draw costs an expected $O(iN)$ by Lemma~\ref{lem:cost}, stopping there costs an expected $O(s^2 N) = O(k^2 N)$ operations for $k \ge 1$ (Corollaries~\ref{cor:sp-sampler}, \ref{cor:u-sampler}, \ref{cor:euclid-sampler}, and~\ref{cor:even-sampler}).

\section{Symplectic Form}\label{sec:symp}

Throughout this section $q$ is arbitrary, $V = \F_q^{2n}$ with the symplectic form, and $G = \Sp(2n,q)$.

\subsection{Orbits}

\begin{theorem}[Classification of alternating forms~{\cite[p.~69]{Taylor}}]\label{thm:alt-classification}
Every non-degenerate alternating form over a field has a symplectic basis; in particular its dimension is even, and two such forms of the same dimension are isometric.
\end{theorem}

Together with Lemma~\ref{lem:transitive} this gives a single orbit.

\begin{theorem}[Orbits, symplectic form]\label{thm:sp-orbit}
$\Cset^{\mathrm{S}}_{n,k,\ell} \ne \emptyset$ if and only if $0 \le \ell \le k$, $k - \ell$ is even, and $k + \ell \le 2n$.
In that case $\Sp(2n,q)$ acts transitively on $\Cset^{\mathrm{S}}_{n,k,\ell}$.
\end{theorem}
\begin{proof}
The conditions are necessary by Lemma~\ref{lem:basic}(2)--(3), and sufficient by Proposition~\ref{prop:sp-rep} below.
For $C, C' \in \Cset^{\mathrm{S}}_{n,k,\ell}$, the non-degenerate quotients are non-degenerate alternating of dimension $k - \ell$, hence isometric by Theorem~\ref{thm:alt-classification}, and Lemma~\ref{lem:transitive} gives transitivity.
\end{proof}

The size of this orbit is the symplectic Anzahl number. The sampler does not need it, but Section~\ref{sec:even} uses it.

\begin{theorem}[Symplectic count~{\cite[Thm.~3.18]{Wan}}]\label{thm:sp-count}
Let $0 \le \ell \le k$ with $k - \ell$ even and $k + \ell \le 2n$.
Then
\begin{equation}\label{eq:sp-count}
|\Cset^{\mathrm{S}}_{n,k,\ell}| \;=\; q^{(k-\ell)(2n-k-\ell)/2}\, \frac{\prod_{i=n-(k+\ell)/2+1}^{n} (q^{2i}-1)}{\prod_{i=1}^{(k-\ell)/2} (q^{2i}-1) \cdot \prod_{i=1}^{\ell} (q^{i}-1)}.
\end{equation}
\end{theorem}

\subsection{Standard representative}

\begin{proposition}[Symplectic representative]\label{prop:sp-rep}
For admissible $(n,k,\ell)$ put $m := (k-\ell)/2$.
The matrix
\[
G_0^{\mathrm{S}} \;=\; \left[\begin{array}{ccc|ccc}
I_\ell & 0 & 0 & 0 & 0 & 0 \\
0 & I_m & 0 & 0 & 0 & 0 \\
0 & 0 & 0 & 0 & I_m & 0
\end{array}\right]
\]
in coordinates $(a \mid b)$, with column blocks of widths $\ell, m, n - \ell - m$ on each side, generates a code in $\Cset^{\mathrm{S}}_{n,k,\ell}$ whose hull is spanned by the first $\ell$ rows.
\end{proposition}
\begin{proof}
The blocks fit, since $\ell + m \le n$ is equivalent to $k + \ell \le 2n$, and the $k$ rows are distinct standard basis vectors, hence independent.
Write $r_1, \ldots, r_k$ for the rows.
For $i \le \ell$, the row $r_i$ has its only non-zero entry in column $i$ of the $a$-part, and column $i$ of the $b$-part is zero in every row, so $r_i$ is orthogonal to all rows.
For $1 \le j \le m$, the rows $r_{\ell+j}$ and $r_{\ell+m+j}$ have their non-zero entries in column $\ell+j$ of the $a$-part and of the $b$-part, respectively, so $B(r_{\ell+j}, r_{\ell+m+j}) = 1$, and all other pairs among these rows are orthogonal.
The Gram matrix is therefore $0_\ell \oplus \bigl(\begin{smallmatrix} 0 & I_m \\ -I_m & 0 \end{smallmatrix}\bigr)$, of rank $2m = k - \ell$; Lemma~\ref{lem:basic}(1) gives hull dimension $\ell$, and the hull is spanned by $r_1, \ldots, r_\ell$.
\end{proof}

\subsection{Uniform sampling of the symplectic group}\label{ssec:sp-sampling}

A uniform element of $\Sp(2n,q)$ is obtained by drawing a uniform symplectic basis.

\begin{proposition}[Uniform symplectic transformation]\label{prop:sp-exact}
Draw vectors $e_1, f_1, \ldots, e_n, f_n$ of $\F_q^{2n}$ successively: with $U_i := \langle e_1, f_1, \ldots, e_{i-1}, f_{i-1}\rangle^\perp$, let $e_i$ be uniform on $U_i \setminus \{0\}$ and $f_i$ uniform on $\{ x \in U_i : B(e_i, x) = 1\}$.
Then the map $g$ with $b_i g = e_i$ and $b_{n+i} g = f_i$ is exactly uniform on $\Sp(2n,q)$, and it is obtained in an expected $O(n^3)$ operations in $\F_q$.
\end{proposition}
\begin{proof}
\emph{Uniformity.}
We apply Lemma~\ref{lem:seq} to the basis of Section~\ref{ssec:codes} in the order $b_1$, $b_{n+1}$, $b_2$, $b_{n+2}$, \ldots, $b_n$, $b_{2n}$, so that $e_i$ and $f_i$ are the images of $b_i$ and $b_{n+i}$.
In this basis $B(b_i, b_{n+i}) = 1$ and all other pairs are orthogonal, and $B(x,x) = 0$ for all $x$.
So the conditions of Lemma~\ref{lem:seq} say that $e_i$ is orthogonal to $e_1, f_1, \ldots, e_{i-1}, f_{i-1}$, and that $f_i$ is orthogonal to these vectors with $B(e_i, f_i) = 1$.
By induction the span of $e_1, f_1, \ldots, e_{i-1}, f_{i-1}$ is non-degenerate, so it meets $U_i$ only in $0$, and
\[
\Sigma_{2i-1} = U_i \setminus \{0\}, \qquad \Sigma_{2i} = \{ x \in U_i : B(e_i, x) = 1 \}
\]
By Lemma~\ref{lem:seq}, $g$ is therefore uniform on $\Sp(2n,q)$ once each $e_i$ and $f_i$ has the stated distribution.
We draw $e_i$ by drawing a uniform element of $U_i$ and redrawing the zero vector, which gives a uniform non-zero vector of $U_i$.
For $f_i$, draw $x$ uniformly from $U_i$ until $t := B(e_i,x) \ne 0$ and put $f_i := t^{-1}x$.
Every $y$ with $B(e_i,y) = 1$ equals $t^{-1}x$ for exactly the $q - 1$ vectors $x = sy$, $s \in \F_q^*$, so $f_i$ is uniform on the required set.

\emph{Cost.}
The subspace $U_i$ is the solution space of the linear system $B(w,x) = 0$ for the vectors $w$ drawn so far, so by Lemma~\ref{lem:cost} it suffices that $e_i$ and $f_i$ need an expected $O(1)$ uniform elements of $U_i$ each.
The zero vector has probability $q^{-2(n-i+1)} \le 1/4$.
Since $U_i$ is non-degenerate and $e_i \ne 0$, $x \mapsto B(e_i,x)$ is a non-zero linear functional on $U_i$, so $t = 0$ has probability $1/q \le 1/2$.
Both loops therefore end after at most two rounds in expectation, so the $i$-th pair costs an expected $O(in)$, and summing over $i \le n$ gives $O(n^3)$.
\end{proof}

For $q = 2$ the samplers of~\cite{KS14,BM21} can be used instead.
The sampler of Koenig and Smolin~\cite{KS14} also builds $g$ from successive symplectic pairs, as above, at the same cost $O(n^3)$; that of Bravyi and Maslov~\cite{BM21} samples a canonical decomposition of the Clifford group in $O(n^2)$ time with the least possible number of random bits.
Both are specific to $q = 2$, while Proposition~\ref{prop:sp-exact} applies to every $q$.

\begin{corollary}[Symplectic sampler]\label{cor:sp-sampler}
For every $q$ and every admissible $(n,k,\ell)$, Algorithm~\ref{alg:main} with Theorem~\ref{thm:sp-orbit}, Proposition~\ref{prop:sp-rep}, and Proposition~\ref{prop:sp-exact} returns an exactly uniform element of $\Cset^{\mathrm{S}}_{n,k,\ell}$ in an expected $O(k^2 n)$ operations in $\F_q$.
\end{corollary}
\begin{proof}
Uniformity follows from Theorem~\ref{thm:uniformity} and Proposition~\ref{prop:sp-exact}.
The rows of $G_0^{\mathrm{S}}$ are $b_i$ $(i \le \ell + m)$ and $b_{n+\ell+j}$ $(j \le m)$, with $\ell + m = (k+\ell)/2$; they are among the images of the first $(k+\ell)/2$ pairs.
By Proposition~\ref{prop:trunc} the sampler of Proposition~\ref{prop:sp-exact} can be stopped after these pairs, which by its proof cost an expected $\sum_{i \le (k+\ell)/2} O(in) = O(k^2 n)$ operations.
\end{proof}

\section{Hermitian Form}\label{sec:herm}

Throughout this section $q$ is arbitrary, $V = \F_{q^2}^n$ with the Hermitian form, $G = \U(n,q^2)$, and $\Nm$ is the norm of Section~\ref{ssec:ff}.

\subsection{Orbits}

\begin{theorem}[Hermitian forms~{\cite[p.~117, Lemma~10.1(v)]{Taylor}}]\label{thm:herm-classification}
Every non-degenerate Hermitian form over $\F_{q^2}$ has an orthonormal basis; in particular two such forms of the same dimension are isometric.
\end{theorem}

Indeed, such a form has an orthogonal basis of non-isotropic vectors~\cite[p.~117]{Taylor}, and as the norm is onto (Section~\ref{ssec:ff}), these can be rescaled to unit vectors.
Together with Lemma~\ref{lem:transitive} this gives a single orbit.

\begin{theorem}[Orbits, Hermitian form]\label{thm:u-orbit}
$\Cset^{\mathrm{H}}_{n,k,\ell} \ne \emptyset$ if and only if $0 \le \ell \le k$ and $k + \ell \le n$.
In that case $\U(n,q^2)$ acts transitively on $\Cset^{\mathrm{H}}_{n,k,\ell}$.
\end{theorem}
\begin{proof}
Necessity is Lemma~\ref{lem:basic}(3) and sufficiency is Proposition~\ref{prop:u-rep}.
The non-degenerate quotients have the same dimension $k - \ell$, hence are isometric by Theorem~\ref{thm:herm-classification}, and Lemma~\ref{lem:transitive} gives transitivity.
\end{proof}

\subsection{Standard representative}

\begin{proposition}[Hermitian representative]\label{prop:u-rep}
Fix $\alpha \in \F_{q^2}$ with $\Nm(\alpha) = -1$: for $q$ even take $\alpha := 1$, and for $q$ odd take $\alpha := \zeta^{(q-1)/2}$ for any non-square $\zeta$ of $\F_{q^2}$, so that $\Nm(\alpha) = \zeta^{(q^2-1)/2} = -1$.
For admissible $(n,k,\ell)$ the matrix
\[
G_0^{\mathrm{H}} \;=\; \bigl[\, \mathrm{diag}\bigl(\underbrace{(1\ \alpha), \ldots, (1\ \alpha)}_{\ell \text{ copies}},\, I_{k-\ell}\bigr) \;\big|\; 0 \,\bigr] \in \F_{q^2}^{k \times n}
\]
generates a code in $\Cset^{\mathrm{H}}_{n,k,\ell}$.
\end{proposition}
\begin{proof}
The first $\ell$ rows of $G_0^{\mathrm{H}}$ have the entries $(1\ \alpha)$ in the disjoint column pairs $\{2i-1, 2i\}$, and the last $k - \ell$ rows are the unit vectors in columns $2\ell+1, \ldots, k+\ell \le n$.
The supports are disjoint, so the rows are independent and pairwise orthogonal.
Each of the first $\ell$ rows $x$ has $B(x,x) = 1 + \Nm(\alpha) = 0$, and each of the others has $B(x,x) = 1$.
The Gram matrix is $0_\ell \oplus I_{k-\ell}$, and Lemma~\ref{lem:basic}(1) gives hull dimension $\ell$.
\end{proof}

\subsection{Uniform sampling of the unitary group}

A non-degenerate Hermitian space of dimension $m \ge 1$ contains
\begin{equation}\label{eq:herm-iso}
Z_m \;:=\; q^{2m-1} + (-1)^m (q^m - q^{m-1})
\end{equation}
isotropic vectors, the zero vector included~\cite[Lemma~10.4]{Taylor}; we use this to bound the rejection rate below.

\begin{proposition}[Uniform unitary transformation]\label{prop:u-exact}
Draw $w_1, \ldots, w_n \in \F_{q^2}^n$ successively: with $U_i := \langle w_1, \ldots, w_{i-1}\rangle^\perp$, let $x$ be uniform on the anisotropic vectors of $U_i$, let $\mu$ be uniform on $\{ \mu : \Nm(\mu) = B(x,x)^{-1}\}$, and put $w_i := \mu x$.
Then the map $g$ with $b_i g = w_i$ is exactly uniform on $\U(n,q^2)$, and it is obtained in an expected $O(n^3 + n \log q)$ operations in $\F_{q^2}$.
\end{proposition}
\begin{proof}
\emph{Uniformity.}
We apply Lemma~\ref{lem:seq} to the standard basis $b_1, \ldots, b_n$, which is orthonormal: $B(b_j, b_i) = 0$ for $j \ne i$ and $B(b_i,b_i) = 1$.
So the conditions of Lemma~\ref{lem:seq} say that $w_i$ is orthogonal to $w_1, \ldots, w_{i-1}$ and that $B(w_i,w_i) = 1$.
Since $\langle w_1, \ldots, w_{i-1}\rangle$ is non-degenerate, it meets $U_i$ only in $0$, so $\Sigma_i$ is the set of unit vectors of $U_i$, and it suffices that $w_i$ is uniform on this set.
Let $A$ be the set of anisotropic vectors of $U_i$; drawing uniformly from $U_i$ until the vector is anisotropic gives $x$ uniform on $A$.
For $\mu$, take any solution $\mu_0$ of $\Nm(\mu_0) = B(x,x)^{-1}$ and put $\mu := \mu_0 \xi^{q-1}$ with $\xi$ uniform on $\F_{q^2}^*$; since $\xi \mapsto \xi^{q-1}$ maps onto $\ker \Nm$ and every element of $\ker\Nm$ is the image of exactly $q - 1$ elements $\xi$ (Section~\ref{ssec:ff}), $\mu$ is uniform on the $q+1$ solutions.
A unit vector $w$ arises exactly from the pairs $(x,\mu) = (\xi w, \xi^{-1})$, $\xi \in \F_{q^2}^*$, each of probability $|A|^{-1}(q+1)^{-1}$, so every unit vector has the same probability.

\emph{Cost.}
A round of the rejection for $x$ is accepted with probability at least $3/8$ by~\eqref{eq:herm-iso}, so by Lemma~\ref{lem:cost} all $x$ together cost an expected $O(n^3)$.
One solution $\mu_0$ of a norm equation is computed for each $i$, at an expected $O(\log q)$ operations (for $q$ even $\mu_0 = c^{q/2}$; for $q$ odd, with $\F_{q^2} = \F_q(\sqrt{d})$, draw $b$ until $c + db^2$ is a square and take $\mu_0 = \sqrt{c+db^2} + b\sqrt{d}$ by Lemma~\ref{lem:cipolla}), and $\xi^{q-1}$ costs $O(\log q)$.
\end{proof}

\begin{corollary}[Hermitian sampler]\label{cor:u-sampler}
For every $q$ and every admissible $(n,k,\ell)$, Algorithm~\ref{alg:main} with Theorem~\ref{thm:u-orbit}, Proposition~\ref{prop:u-rep}, and Proposition~\ref{prop:u-exact} returns an exactly uniform element of $\Cset^{\mathrm{H}}_{n,k,\ell}$ in an expected $O(k^2 n + k \log q)$ operations in $\F_{q^2}$.
\end{corollary}
\begin{proof}
Uniformity follows from Theorem~\ref{thm:uniformity} and Proposition~\ref{prop:u-exact}.
Since $G_0^{\mathrm{H}}$ uses only its first $k+\ell$ columns, Proposition~\ref{prop:trunc} stops the sampler of Proposition~\ref{prop:u-exact} after $w_{k+\ell}$; by its proof this costs an expected $\sum_{i \le k+\ell} O(in) = O(k^2 n)$ operations and $k + \ell \le 2k$ solutions of norm equations, each costing an expected $O(\log q)$.
\end{proof}

\section{Euclidean Form, $q$ Odd}\label{sec:euclid-odd}

Throughout this section $q$ is odd, $V = \F_q^n$ with the Euclidean form, and $G = \OO(n,q)$.

\subsection{Orbits}\label{ssec:euclid-orbits}

The \emph{discriminant} of a non-degenerate symmetric form is the determinant of a Gram matrix modulo $(\F_q^*)^2$, an element of $\F_q^*/(\F_q^*)^2 \cong \{\pm 1\}$ that does not depend on the basis; the standard form has Gram matrix $I_n$ and discriminant $1$.
The \emph{Witt type} of a code $C$ is the discriminant of its non-degenerate quotient,
\[
\epsilon(C) \;:=\; \disc(\bar B_C) \;\in\; \{\pm 1\},
\]
with $\epsilon(C) := 1$ when $B|_C = 0$; equivalently, $\epsilon(C) = \disc(B|_W)$ for any complement $W$ of $\Rad(C)$ in $C$.
We write $\Cset_{n,k,\ell}^\epsilon$ for the codes of Witt type $\epsilon$; in the notation of~\cite{LiShiLing2025}, $\Cset_{n,k,\ell}^{+1}$ and $\Cset_{n,k,\ell}^{-1}$ are the sets $\mathrm{Hull}^\ell_+[n,k]_q$ and $\mathrm{Hull}^\ell_-[n,k]_q$ of codes of Type~$\mathcal{S}$ and Type~$\mathcal{NS}$, since the Type is defined by whether the discriminant of $\bar B_C$ is a square~\cite[Prop.~3]{LiShiLing2025}.

\begin{theorem}[Classification of symmetric forms~{\cite[Lemma~6]{MacWilliams69}}]\label{thm:disc-classification}
Two non-degenerate symmetric bilinear forms over $\F_q$ of the same dimension are isometric if and only if they have the same discriminant.
\end{theorem}

\begin{lemma}[Witt type is invariant]\label{lem:witt-invariant}
For $g \in \OO(n,q)$ and every code $C$, $\epsilon(Cg) = \epsilon(C)$.
\end{lemma}
\begin{proof}
Since $g$ maps $\Rad(C)$ onto $\Rad(Cg)$, it induces a linear bijection $C/\Rad(C) \to Cg/\Rad(Cg)$, and $\bar B_{Cg}(cg + \Rad(Cg), c'g + \Rad(Cg)) = B(cg, c'g) = B(c,c')$; so the induced forms are isometric and have the same discriminant.
\end{proof}

\begin{theorem}[Orbits, Euclidean form, $q$ odd]\label{thm:orbits}
Each non-empty $\Cset_{n,k,\ell}^\epsilon$ is a single $\OO(n,q)$-orbit, so $\Cset_{n,k,\ell}$ consists of at most two orbits.
\end{theorem}
\begin{proof}
Different Witt types lie in different orbits by Lemma~\ref{lem:witt-invariant}.
If $\epsilon(C) = \epsilon(C')$, the non-degenerate quotients of $C$ and $C'$ have the same dimension $k - \ell$ and the same discriminant, so they are isometric by Theorem~\ref{thm:disc-classification}, and Lemma~\ref{lem:transitive} gives $g$ with $Cg = C'$.
\end{proof}

The largest dimension of a totally isotropic subspace is the Witt index of the standard form, which is $\lfloor n/2 \rfloor$, except that it is $n/2 - 1$ when $n$ is even and the form has minus type~\cite[pp.~138--139]{Taylor}, that is, $\eta((-1)^{n/2}) = -1$ since the standard form has discriminant $1$; see also~\cite[\S 3.2]{RainsSloane98}. Hence
\[
k_{\max}(n,q) \;=\; \begin{cases} \lfloor n/2 \rfloor - 1 & n \equiv 2 \pmod 4 \text{ and } q \equiv 3 \pmod 4, \\ \lfloor n/2 \rfloor & \text{otherwise.} \end{cases}
\]
The next lemma records exactly which Witt types occur.

\begin{lemma}[Non-empty orbits]\label{lem:nonempty}
$\Cset_{n,k,\ell}^\epsilon \ne \emptyset$ if and only if
\begin{equation}\label{eq:nonempty}
0 \le \ell \le k, \qquad k + \ell \le n, \qquad \ell \le k_{\max}(n,q), \qquad \epsilon \in E(n,k,\ell),
\end{equation}
where $E(n,k,\ell) := \{1\}$ if $k = \ell$, $E(n,k,\ell) := \{\eta((-1)^\ell)\}$ if $n = k + \ell$ and $\ell < k$, and $E(n,k,\ell) := \{\pm 1\}$ otherwise.
\end{lemma}
\begin{proof}
The first two conditions are necessary by Lemma~\ref{lem:basic}, and the third holds because $\Rad(C)$ is totally isotropic.
Let $H := \Rad(C)$ and choose a Witt decomposition $V = H \oplus M \oplus H^*$.
Then $V$ is the orthogonal sum of $M \cong H^\perp/H$ and the space $H \oplus H^*$, whose Gram matrix $\bigl(\begin{smallmatrix} 0 & I_\ell \\ I_\ell & 0 \end{smallmatrix}\bigr)$ has discriminant $\eta((-1)^\ell)$.
Since discriminants multiply on orthogonal sums and $\disc(B) = 1$, writing $H^\perp/H = \bar C \perp \bar C^\perp$ with $\bar C := C/H$ gives
\begin{equation}\label{eq:disc-quotient}
\epsilon(C) \cdot \disc(\bar C^\perp) \;=\; \eta\bigl((-1)^\ell\bigr).
\end{equation}
If $k = \ell$ then $\bar C = 0$ and $\epsilon(C) = 1$; if $n = k + \ell$ then $\bar C^\perp = 0$ and~\eqref{eq:disc-quotient} gives $\epsilon(C) = \eta((-1)^\ell)$.
Conversely, Proposition~\ref{prop:euclid-reps} constructs a code in $\Cset_{n,k,\ell}^\epsilon$ for every tuple satisfying~\eqref{eq:nonempty}.
\end{proof}

The orbit sizes are given by the mass formula of Li, Shi, and Ling.
Write $\eta_j := \eta((-1)^j)$ and $\binom{a}{b}_{Q} := \prod_{i=0}^{b-1} \frac{Q^{a-i}-1}{Q^{i+1}-1}$ for the Gaussian binomial coefficient.

\begin{theorem}[Euclidean count, $q$ odd~{\cite[Thms.~4, 5]{LiShiLing2025}}]\label{thm:orbit-size-odd}
Let $q$ be odd, $\epsilon \in \{\pm 1\}$, and $1 \le k_0 := k - \ell < n$ with $k + \ell \le n$.
For $\ell = 0$~\cite[Thm.~31]{CMTQ19}, $|\Cset_{n,k_0,0}^{\epsilon}|$ equals, in the four cases (i) $k_0$ odd and $n$ even, (ii) $k_0$ and $n$ odd, (iii) $k_0$ even and $n$ odd, (iv) $k_0$ and $n$ even,
\begin{align*}
&\text{(i)} && \tfrac12 q^{(k_0(n-k_0)-1)/2} \bigl(q^{n/2} - \eta_{n/2}\bigr) \tbinom{n/2-1}{(k_0-1)/2}_{q^2},\\
&\text{(ii)} && \tfrac12 q^{k_0(n-k_0)/2} \bigl(q^{(n-k_0)/2} + \epsilon\,\eta_{(n-k_0)/2}\bigr) \tbinom{(n-1)/2}{(k_0-1)/2}_{q^2},\\
&\text{(iii)} && \tfrac12 q^{k_0(n-k_0)/2} \bigl(q^{k_0/2} + \epsilon\,\eta_{k_0/2}\bigr) \tbinom{(n-1)/2}{k_0/2}_{q^2},\\
&\text{(iv)} && \tfrac12 q^{k_0(n-k_0)/2}\, \frac{(q^{k_0/2} + \epsilon\,\eta_{k_0/2})(q^{(n-k_0)/2} + \epsilon\,\eta_{(n-k_0)/2})}{q^{n/2} + \eta_{n/2}} \tbinom{n/2}{k_0/2}_{q^2},
\end{align*}
and for $\ell \ge 1$
\begin{equation}\label{eq:orbit-size-odd}
|\Cset_{n,k,\ell}^{\epsilon}| = \begin{cases}
\displaystyle\prod_{i=1}^{\ell} \frac{q^{n-k_0-2i+1}-1}{q^{k_0}(q^i-1)} \, |\Cset_{n,k_0,0}^{\epsilon}|, & n - k_0 \text{ odd},\\[10pt]
\displaystyle\frac{q^{m-\ell} + \epsilon\,\eta_m}{q^{m} + \epsilon\,\eta_m} \prod_{i=1}^{\ell} \frac{q^{n-k_0-2i+2}-1}{q^{k_0}(q^i-1)} \, |\Cset_{n,k_0,0}^{\epsilon}|, & n - k_0 = 2m \text{ even}.
\end{cases}
\end{equation}
\end{theorem}

For $k = \ell$ there is a single orbit, and for $k_0 = n$ the only code is $V$; in these cases no weights are needed.

\subsection{Standard representatives}

By Lemmas~\ref{lem:minus-one-square} and~\ref{lem:sum-two-squares} we can fix $\theta$ with $\theta^2 = -1$ if $q \equiv 1 \pmod 4$, and $a, b$ with $a^2 + b^2 = -1$ if $q \equiv 3 \pmod 4$, and put
\[
M := \begin{pmatrix} 1 & 0 & a & b \\ 0 & 1 & -b & a \end{pmatrix}, \qquad r := (1\ a\ b),
\]
so that $MM^\top = 0$ and $rr^\top = 0$.
Fix also $a', b'$ with $a'^2 + b'^2$ a non-square (take $(a, b)$ if $q \equiv 3 \pmod 4$).
We distinguish three cases:
(a) $q \equiv 1 \pmod 4$;
(b) $q \equiv 3 \pmod 4$ and $\ell$ is even;
(c) $q \equiv 3 \pmod 4$ and $\ell$ is odd.
The representatives are built from a totally isotropic block for the radical and a non-degenerate block for a complement:
\begin{itemize}
\item \emph{hull block} $G_H \in \F_q^{\ell \times \ell_H}$:
$G_H := (I_\ell \mid \theta I_\ell)$ in case (a);
$G_H := \mathrm{diag}(M, \ldots, M)$ in case (b);
$G_H := \mathrm{diag}(M, \ldots, M, r)$ in case (c).
Here $\ell_H = 2\ell$ in (a) and (b) and $\ell_H = 2\ell + 1$ in (c), and $G_H G_H^\top = 0$.
\item \emph{LCD block} $G_W^\epsilon$, with $m := k - \ell$: $G_W^+ := I_m$, and $G_W^- := \mathrm{diag}(I_{m-1}, (a'\ b')) \in \F_q^{m \times (m+1)}$, so that $G_W^\epsilon (G_W^\epsilon)^\top$ is diagonal with discriminant $\epsilon$.
\end{itemize}

\begin{proposition}[Euclidean representatives, $q$ odd]\label{prop:euclid-reps}
Let $(n,k,\ell,\epsilon)$ satisfy~\eqref{eq:nonempty}.
\begin{enumerate}
\item Except in case (c) with $\epsilon = -1$, the matrix $G_0^\epsilon := [\,\mathrm{diag}(G_H, G_W^\epsilon) \mid 0\,] \in \F_q^{k \times n}$, padded with zero columns, generates a code in $\Cset_{n,k,\ell}^\epsilon$.
\item In case (c) with $\epsilon = -1$, the matrix
\[
G_0^- \;:=\; \left[\;
\begin{array}{c|c|c|c}
\mathrm{diag}(M, \ldots, M) & 0 & 0 & 0 \\ \hline
0 & T & 0 & 0 \\ \hline
0 & 0 & I_{k-\ell-1} & 0
\end{array}
\;\right],
\]
with $(\ell-1)/2$ copies of $M$ and $T := \bigl(\begin{smallmatrix} 1 & a & b \\ 0 & -b & a \end{smallmatrix}\bigr)$, generates a code in $\Cset_{n,k,\ell}^-$.
\end{enumerate}
\end{proposition}
\begin{proof}
(1) The rows are independent, and the Gram matrix is $G_H G_H^\top \oplus G_W^\epsilon (G_W^\epsilon)^\top = 0_\ell \oplus D$ with $D$ diagonal, non-singular, of discriminant $\epsilon$.
By Lemma~\ref{lem:basic}(1) the hull dimension is $\ell$; the radical is the row space of $G_H$ and the rows of $G_W^\epsilon$ span a complement of discriminant $\epsilon$.
It remains to check that the blocks fit in $n$ columns.
Their width is $k + \ell$ in cases (a) and (b) with $\epsilon = +1$, and $k + \ell + 1$ in cases (a) and (b) with $\epsilon = -1$ and in case (c) with $\epsilon = +1$.
The first is at most $n$ by~\eqref{eq:nonempty}.
For the second we need $n \ne k + \ell$.
If $\ell < k$ and $n = k + \ell$, then $\epsilon = \eta((-1)^\ell)$ by Lemma~\ref{lem:nonempty}, which is $+1$ in cases (a) and (b) and $-1$ in case (c), opposite to the cases at hand.
If $\ell = k$, only $\epsilon = +1$ occurs, and $n = k + \ell = 2\ell$ in case (c) would give $n \equiv 2 \pmod 4$ with $q \equiv 3 \pmod 4$ and $\ell = n/2 > k_{\max}(n,q)$.

(2) The construction of (1) would need $k + \ell + 2$ columns, two more than may be available.
Instead, the last hull row $r = (1\ a\ b)$ and the first LCD row $v := (0\ {-b}\ a)$ share three columns in the block $T$; they are orthogonal, since $B(v,r) = -ba + ab = 0$, and $B(v,v) = a^2 + b^2 = -1$ is a non-square.
The Gram matrix is $0_\ell \oplus \mathrm{diag}(-1, 1, \ldots, 1)$, so the hull dimension is $\ell$ and the Witt type is $\eta(-1) = -1$.
The last non-zero column is $k + \ell \le n$.
\end{proof}

\subsection{Uniform sampling of the orthogonal group}\label{ssec:euclid-sampling}

As for the unitary group, we draw an orthonormal basis by rejection and rescaling, which now needs a square root.
Squares are recognized by Euler's criterion and square roots are computed by Lemma~\ref{lem:cipolla}.

The rejection rate of the orthogonal sampler is bounded with the numbers of solutions of diagonal quadratic equations.

\begin{lemma}[Solution counts~{\cite[Thms.~6.26--6.27]{LidlNiederreiter}}]\label{lem:nsol}
Let $a_1, \ldots, a_d \in \F_q^*$, $\Delta := a_1 \cdots a_d$, and $c \in \F_q$.
The number $N_d(a;c)$ of $x \in \F_q^d$ with $\sum_j a_j x_j^2 = c$ is $q^{d-1} + v(c)\, q^{d/2 - 1}\, \eta((-1)^{d/2}\Delta)$ for $d$ even, where $v(0) = q-1$ and $v(c) = -1$ for $c \ne 0$, and $q^{d-1} + q^{(d-1)/2}\, \eta((-1)^{(d-1)/2} c \Delta)$ for $d$ odd.
\end{lemma}

\begin{proposition}[Uniform orthogonal transformation]\label{prop:o-exact}
Draw $w_1, \ldots, w_n$ successively: with $U_i := \langle w_1, \ldots, w_{i-1}\rangle^\perp$, draw $x$ uniformly from $U_i$ until $B(x,x)$ is a non-zero square, compute a square root $s$ of $B(x,x)$ by Lemma~\ref{lem:cipolla}, and put $w_i := x / s$.
Then the map $g$ with $b_i g = w_i$ is exactly uniform on $\OO(n,q)$, and it is obtained in an expected $O(n^3 + n \log q)$ operations in $\F_q$.
\end{proposition}
\begin{proof}
\emph{Uniformity.}
We apply Lemma~\ref{lem:seq} to the standard basis $b_1, \ldots, b_n$, which is orthonormal; as in the proof of Proposition~\ref{prop:u-exact}, $\Sigma_i$ is the set of unit vectors of $U_i$, and it suffices that $w_i$ is uniform on this set.
Let $A$ be the set of $x \in U_i$ with $B(x,x)$ a non-zero square, so that $x$ is uniform on $A$, and let $\pi_c(s)$ be the probability that the square-root computation returns $s$ on input $c$; it depends only on $c$, and $\pi_c(s) + \pi_c(-s) = 1$ for $s^2 = c$.
A unit vector $w$ arises exactly from the pairs $(x, s) = (sw, s)$ with $s \in \F_q^*$, so it is returned with probability $|A|^{-1} \sum_{s \in \F_q^*} \pi_{s^2}(s) = |A|^{-1} (q-1)/2$, the same for every unit vector.

\emph{Cost.}
We bound the success probability of a round.
The space $U_i$ is non-degenerate of dimension $d := n - i + 1$ and contains a unit vector, since $\Sigma_i \ne \emptyset$.
If $d = 1$, then $U_i = \langle w \rangle$ with $B(w,w) = 1$, and $B(x,x)$ is a non-zero square unless $x = 0$, with probability $1 - 1/q \ge 2/3$.
If $d \ge 2$, Lemma~\ref{lem:nsol} in an orthogonal basis of $U_i$ shows that each $c \ne 0$ is taken by at least $q^{d-1} - q^{d/2-1} \ge (q-1)q^{d-2}$ vectors for $d$ even and at least $q^{d-1} - q^{(d-1)/2} \ge (q-1)q^{d-2}$ for $d \ge 3$ odd, so a round succeeds with probability at least $\frac{q-1}{2} \cdot \frac{q-1}{q^2} \ge \frac29$.
By Lemma~\ref{lem:cost} the rounds cost an expected $O(n^3)$ operations in total.
Each round also evaluates $\eta$ once, at $O(\log q)$ operations, and each vector needs one square root, at an expected $O(\log q)$ operations by Lemma~\ref{lem:cipolla}, which adds $O(n \log q)$.
\end{proof}

\begin{corollary}[Euclidean sampler, $q$ odd]\label{cor:euclid-sampler}
For every admissible $(n,k,\ell)$, Algorithm~\ref{alg:main} with the orbits of Theorem~\ref{thm:orbits}, the weights $p^\epsilon$ from~\eqref{eq:orbit-size-odd} (with $p^\epsilon := 0$ for $\epsilon \notin E(n,k,\ell)$), the representatives of Proposition~\ref{prop:euclid-reps}, and the group sampler of Proposition~\ref{prop:o-exact} returns an exactly uniform element of $\Cset^{\mathrm{E}}_{n,k,\ell}$ in an expected $O(k^2 n + k \log q)$ operations in $\F_q$.
\end{corollary}
\begin{proof}
Uniformity follows from Theorem~\ref{thm:uniformity}, Theorem~\ref{thm:orbits}, and Proposition~\ref{prop:o-exact}.
Since $G_0^\epsilon$ uses only its first $k + \ell + 1$ columns (proof of Proposition~\ref{prop:euclid-reps}), Proposition~\ref{prop:trunc} stops the sampler after $w_{k+\ell+1}$; by its proof this costs an expected $\sum_{i \le k+\ell+1} O(in) + O((k + \ell + 1)\log q) = O(k^2 n + k \log q)$ operations.
The weights are computed once from Theorem~\ref{thm:orbit-size-odd}: the factors that do not involve $\epsilon$ are common to both signs, and the factor $q^m + \epsilon\,\eta_m$ of~\eqref{eq:orbit-size-odd} cancels against the same factor of $|\Cset^\epsilon_{n,k_0,0}|$, so $p^+ : p^-$ is a ratio of two products of at most two factors of the form $q^j \pm 1$.
\end{proof}

\section{Euclidean Form, $q$ Even}\label{sec:even}

Throughout this section $q$ is even, $V = \F_q^n$ with the Euclidean form, and $G = \Ps(n,q)$ is defined below.
Put $\lambda(x) := \sum_i x_i$, so that $B(x,x) = \lambda(x)^2$, and $\mathbf{1} := (1, \ldots, 1)$; then $B(\mathbf{1},y) = \lambda(y)$.
The form is symmetric with Gram matrix $I_n$ but not alternating: it vanishes on the diagonal exactly on the hyperplane $V_0 := \ker\lambda = \mathbf{1}^\perp$, the even-weight subspace when $q = 2$.
Theorem~\ref{thm:witt} is therefore not available, and the isometry group
\[
\Ps(n,q) \;:=\; \{ g \in \GL(V) : B(xg,yg) = B(x,y) \ \forall x,y \}
\]
is the \emph{pseudo-symplectic group}~\cite{LiuWan}.
We use two known results: Theorem~\ref{thm:ps-group} below, due to Dieudonn\'e~\cite{Dieudonne48} (see~\cite[p.~370]{Wan97}), and Theorem~\ref{thm:ps-orbits} in Section~\ref{ssec:ps-orbits}, due to Liu and Wan~\cite{LiuWan} (see also~\cite[(19)--(25)]{Wan97}).
For completeness, we prove them in Appendix~\ref{app:ps}.
The first reduces $\Ps(n,q)$ to a symplectic group.

\newcommand{\PsGroupStatement}{\begin{enumerate}
\item Every element of $\Ps(n,q)$ fixes $\mathbf{1}$ and preserves $V_0$.
\item If $n$ is odd, then $V = \langle \mathbf{1}\rangle \perp V_0$ with $V_0$ symplectic of dimension $n-1$, and $g \mapsto g|_{V_0}$ is an isomorphism $\Ps(n,q) \to \Sp(V_0)$, so $\Ps(n,q) \cong \Sp(n-1,q)$.
\item If $n$ is even, then for every $u$ with $\lambda(u) = 1$, $V = \langle \mathbf{1}, u\rangle \perp \langle \mathbf{1}, u\rangle^\perp$ with $\langle \mathbf{1}, u\rangle^\perp$ symplectic of dimension $n-2$.
For any $u, u'$ with $\lambda(u) = \lambda(u') = 1$, the map $g \mapsto g|_{\langle \mathbf{1}, u\rangle^\perp}$ is a bijection from $\{ g \in \Ps(n,q) : ug = u' \}$ onto the symplectic isomorphisms $\langle \mathbf{1}, u\rangle^\perp \to \langle \mathbf{1}, u'\rangle^\perp$.
Hence $\Ps(n,q)$ acts transitively on the $q^{n-1}$ vectors $u$ with $\lambda(u) = 1$, with stabilizers isomorphic to $\Sp(n-2,q)$, and $|\Ps(n,q)| = q^{n-1} |\Sp(n-2,q)|$.
\end{enumerate}}
\begin{theorem}[The group~{\cite{Dieudonne48}, cf.~\cite{MacWilliams69}}]\label{thm:ps-group}
\PsGroupStatement
\end{theorem}

So the geometry of $V$ reduces to that of the symplectic space $V_0$ for $n$ odd, and of $\langle \mathbf{1}, u\rangle^\perp$ for $n$ even.

\subsection{Orbits}\label{ssec:ps-orbits}

For a code $C$ put $\tau(C) := 0$ if $C \subseteq V_0$ and $\tau(C) := 1$ otherwise, and $\iota(C) := 1$ if $\mathbf{1} \in C$ and $\iota(C) := 0$ otherwise.
Thus $\tau(C) = 0$ says that $B|_C$ is alternating.

Let $\nu := \lfloor (n-1)/2 \rfloor$, so that $2\nu = n - 1$ for $n$ odd and $2\nu = n - 2$ for $n$ even.

\newcommand{\PsOrbitsStatement}{Let $1 \le k \le n$ and $0 \le \ell \le k$.
\begin{enumerate}
\item Two codes of $\Cset_{n,k,\ell}$ lie in the same $\Ps(n,q)$-orbit if and only if they have the same pair $(\tau,\iota)$.
Hence $\Cset_{n,k,\ell}$ consists of at most four orbits $\Cset_{n,k,\ell}^{\tau,\iota}$.
\item $\Cset_{n,k,\ell}^{\tau,\iota}$ is non-empty if and only if $(n \bmod 2, \tau, \iota)$ is listed in Table~\ref{tab:ps-orbits} and the condition there holds, and then its size is the one listed there.
\end{enumerate}}
\newcommand{\PsOrbitsTabular}{\setlength{\tabcolsep}{4pt}
\begin{tabular}{ccll}
\hline
$n$ & $(\tau,\iota)$ & condition & size \\
\hline
odd & $(0,0)$ & $k-\ell$ even, $k+\ell \le n-1$ & $|\Cset^{\mathrm{S}}_{\nu,k,\ell}|$ \\
odd & $(1,0)$ & $\ell < k$, $k+\ell \le n-1$ & $(q^{k-\ell}-1)\,|\Cset^{\mathrm{S}}_{\nu,k,\ell}|$ \\
 & & & ${}+ (q^{k}-q^{k-\ell-1})\,|\Cset^{\mathrm{S}}_{\nu,k,\ell+1}|$ \\
odd & $(1,1)$ & $k-\ell$ odd, $k+\ell \le n$ & $|\Cset^{\mathrm{S}}_{\nu,k-1,\ell}|$ \\
even & $(0,0)$ & $k-\ell$ even, $k+\ell \le n-1$ & $q^{k}\,|\Cset^{\mathrm{S}}_{\nu,k,\ell}|$ \\
even & $(1,0)$ & $\ell < k$, $k+\ell \le n-1$ & $q^{\,n-1-\ell}\,|\Cset^{\mathrm{S}}_{\nu,k-1,\ell}|$ \\
 & & & ${}+ \bigl(q^{\,n-1}-q^{\,n-2-\ell}\bigr)\,|\Cset^{\mathrm{S}}_{\nu,k-1,\ell+1}|$ \\
even & $(1,1)$ & $k-\ell \ge 2$ even, $k+\ell \le n$ & $q^{\,n-k}\,|\Cset^{\mathrm{S}}_{\nu,k-2,\ell}|$ \\
even & $(0,1)$ & $\ell \ge 1$, $k-\ell$ even, $k+\ell \le n$ & $|\Cset^{\mathrm{S}}_{\nu,k-1,\ell-1}|$ \\
\hline
\end{tabular}}
\begin{theorem}[Orbits, Euclidean form, $q$ even~{\cite{LiuWan}, \cite[(22)--(25)]{Wan97}}]\label{thm:ps-orbits}
\PsOrbitsStatement
\end{theorem}

\begin{table}[h]
\centering
\caption{The non-empty $\Ps(n,q)$-orbits $\Cset_{n,k,\ell}^{\tau,\iota}$ for $q$ even, with $\nu = \lfloor (n-1)/2 \rfloor$; for $n$ odd, $\Cset_{n,k,\ell}^{0,1}$ is empty.}\label{tab:ps-orbits}
\medskip
\PsOrbitsTabular
\end{table}

\begin{remark}
In the notation of Liu and Wan~\cite{LiuWan} (see~\cite[pp.~370--371]{Wan97}), $\Cset_{n,k,\ell}^{\tau,\iota}$ is the set of subspaces of type $(k,\ k-\ell,\ s,\ \iota)$ with $2s + z = k - \ell$, where $z = 0$ if $\tau = 0$, $z = 1$ if $\tau = 1$ and $k - \ell$ is odd, and $z = 2$ if $\tau = 1$ and $k - \ell$ is even.
In this notation the conditions of Theorem~\ref{thm:ps-orbits}(2) are~\cite[(22)--(23)]{Wan97}, and the sizes are the orbit lengths $N(k, k-\ell, s, \iota; n)$ of~\cite[(25)]{Wan97}, written in terms of the symplectic counts, with $n_0 = 0$ for $n$ odd (the value $n_0 = 1$ printed there for $\delta = 1$ is a misprint).
\end{remark}

\subsection{Standard representatives}

As for $q$ odd, the representatives are assembled from a few small blocks, now with entries in $\{0,1\}$, so the construction is the same for every even $q$:
\begin{itemize}
\item \emph{hull block} $H_\ell := \mathrm{diag}\bigl((1\ 1), \ldots, (1\ 1)\bigr) \in \F_q^{\ell \times 2\ell}$, whose rows $b_{2i-1} + b_{2i}$ are isotropic, of even weight, and pairwise orthogonal;
\item \emph{alternating block} $J_m \in \F_q^{m \times (m+1)}$ for even $m \ge 2$, with rows $b_j + b_{j+1}$ $(1 \le j \le m)$, and $J_0$ the empty matrix;
\item \emph{anisotropic block} $I_m$.
\end{itemize}
The rows of $J_m$ have even weight, and their Gram matrix is the tridiagonal matrix with zero diagonal and ones next to it, the adjacency matrix $A_m$ of a path on $m$ vertices.
Block-diagonal matrices are padded with zero columns to width $n$.

\begin{lemma}[Path matrix]\label{lem:path}
The determinant $d_m$ of $A_m$ satisfies $d_m = -d_{m-2}$ for $m \ge 3$, with $d_1 = 0$ and $d_2 = -1$ (and $d_0 := 1$); so $A_m$ is non-singular for even $m$ and singular for odd $m$.
\end{lemma}
\begin{proof}
Expanding along the first row, the only non-zero entry is the $1$ in position $(1,2)$, and its minor has first column $(1, 0, \ldots, 0)^\top$, so it equals $d_{m-2}$; hence $d_m = -d_{m-2}$.
\end{proof}

\begin{proposition}[Representatives, $q$ even]\label{prop:ps-reps}
If $(n,k,\ell,\tau,\iota)$ satisfies the condition of Theorem~\ref{thm:ps-orbits}(2), then the matrix of Table~\ref{tab:ps-reps} generates a code in $\Cset_{n,k,\ell}^{\tau,\iota}$.
\end{proposition}

\begin{table}[h]
\centering
\caption{Representatives of the $\Ps(n,q)$-orbits for $q$ even. Rows separated by a semicolon are stacked; $\mathrm{diag}(\cdot)$ is padded with zero columns to width $n$, and $\mathrm{diag}(\cdot)^{+}$ is $\mathrm{diag}(\cdot)$ with $\mathbf{1}$ added to its row $\ell+1$.}\label{tab:ps-reps}
\medskip
\begin{tabular}{cccl}
\hline
$n$ & $(\tau,\iota)$ & condition & generator matrix \\
\hline
any & $(0,0)$ & $k-\ell$ even, $k+\ell \le n-1$ & $\mathrm{diag}(H_\ell, J_{k-\ell})$ \\
odd & $(1,0)$ & $k-\ell \ge 2$ even, $k+\ell \le n-1$ & $\mathrm{diag}(H_\ell, J_{k-\ell})^{+}$ \\
odd & $(1,0)$ & $k-\ell$ odd, $k+\ell \le n-1$ & $\mathrm{diag}(H_{\ell+1}, J_{k-\ell-1})^{+}$ \\
even & $(1,0)$ & $k-\ell$ odd, $k+\ell \le n-1$ & $b_n;\ \mathrm{diag}(H_\ell, J_{k-\ell-1})$ \\
even & $(1,0)$ & $k-\ell \ge 2$ even, $k+\ell \le n-1$ & $b_n;\ \mathrm{diag}(H_{\ell+1}, J_{k-\ell-2})^{+}$ \\
odd & $(1,1)$ & $k-\ell$ odd, $k+\ell \le n$ & $\mathbf{1};\ \mathrm{diag}(H_\ell, J_{k-\ell-1})$ \\
even & $(1,1)$ & $k-\ell \ge 2$ even, $k+\ell \le n$ & $\mathbf{1};\ b_n;\ \mathrm{diag}(H_\ell, J_{k-\ell-2})$ \\
even & $(0,1)$ & $\ell \ge 1$, $k-\ell$ even, $k+\ell \le n$ & $\mathbf{1};\ \mathrm{diag}(H_{\ell-1}, J_{k-\ell})$ \\
\hline
\end{tabular}
\end{table}

\begin{proof}
The block $\mathrm{diag}(H_a, J_m)$ occupies $2a + m + 1$ columns if $m \ge 2$ and $2a$ columns if $m = 0$, hence at most $k + \ell + 2$ columns in every row of the table.
Under the conditions of Theorem~\ref{thm:ps-orbits}(2) this is at most $n - 1$ for $n$ even and at most $n$ for $n$ odd, since $k + \ell \equiv k - \ell \pmod 2$ (e.g., $k + \ell \le n - 2$ for $n$ even, $(\tau,\iota) = (1,0)$, and $k - \ell$ even); in particular column $n$ of the block is zero for $n$ even.
We use $B(\mathbf{1},x) = \lambda(x)$, $B(\mathbf{1},\mathbf{1}) = n \bmod 2$, and $B(b_n,b_n) = B(b_n,\mathbf{1}) = 1$; blocks with disjoint supports are orthogonal.
We give the Gram matrix and the invariants in each case.

\emph{Case $(0,0)$.} The Gram matrix is $0_\ell \oplus A$ with $A$ the non-singular path matrix of size $k - \ell$, so $\dim\Hull(C) = \ell$; all rows have even weight, so $\tau = 0$; and $\mathbf{1} \notin C$, since $C \subseteq V_0 \not\ni \mathbf{1}$ for $n$ odd and column $n$ is zero for $n$ even.

\emph{Case $(1,0)$, $n$ odd.} Write $y_1, \ldots, y_k$ for the block rows, which have even weight, so that $B(y_i + \delta\mathbf{1}, y_j + \delta'\mathbf{1}) = B(y_i,y_j) + \delta\delta'$.
The Gram matrix is that of the block plus a $1$ in position $(\ell+1,\ell+1)$: for $k - \ell$ even it is $0_\ell \oplus A'$, where $A'$ is the path matrix with an extra $1$ in its first diagonal entry; expanding along the first row, $\det A' = d_{k-\ell-1} - d_{k-\ell-2} = d_{k-\ell} \ne 0$ by Lemma~\ref{lem:path}, since $d_{k-\ell-1} = 0$; for $k - \ell$ odd it is $0_\ell \oplus (1) \oplus A$.
So $\dim\Hull(C) = \ell$; row $\ell+1$ has $\lambda = 1$, so $\tau = 1$; and $\mathbf{1} \notin C$: if $\mathbf{1} = \sum_i c_i y_i + c\,\mathbf{1}$, where $c$ is the coefficient of row $\ell+1$, then applying $\lambda$, which vanishes on the $y_i$, gives $c = 1$, so $\sum_i c_i y_i = 0$ with $c_{\ell+1} = 1$, contradicting the independence of the block rows; so $\iota = 0$.

\emph{Case $(1,0)$, $n$ even.} The vector $b_n$ is orthogonal to the block rows, $B(b_n, y_{\ell+1} + \mathbf{1}) = 1$, and $y_{\ell+1} + \mathbf{1}$ is isotropic and orthogonal to the other block rows; so the Gram matrix is $(1) \oplus 0_\ell \oplus A$ for $k - \ell$ odd and $\bigl(\begin{smallmatrix}1&1\\1&0\end{smallmatrix}\bigr) \oplus 0_\ell \oplus A$ for $k - \ell$ even, and $\dim\Hull(C) = \ell$.
$\lambda(b_n) = 1$ gives $\tau = 1$.
All rows other than $b_n$ lie in $V_0$, and $\lambda(b_n) = 1$, so $C \cap V_0$ is spanned by them; as $\mathbf{1} \in V_0$, $\mathbf{1} \in C$ would give $\mathbf{1} = \sum_i c_i y_i + c\,\mathbf{1}$, with $c$ the coefficient of row $\ell+1$ ($c = 0$ for $k - \ell$ odd).
Since column $n$ of the block is zero, column $n$ gives $c = 1$, which is impossible for $k - \ell$ odd; for $k - \ell$ even it gives $\sum_i c_i y_i = 0$ with $c_{\ell+1} = 1$, contradicting the independence of the block rows; so $\iota = 0$.

\emph{Case $(1,1)$, $n$ odd.} The Gram matrix is $(1) \oplus 0_\ell \oplus A$, because $\mathbf{1}$ is anisotropic and orthogonal to the even-weight rows, so $\dim\Hull(C) = \ell$; $\lambda(\mathbf{1}) = 1$ gives $\tau = \iota = 1$.
The rows are independent because $\mathbf{1} \notin V_0$.

\emph{Case $(1,1)$, $n$ even.} The Gram matrix is $\bigl(\begin{smallmatrix}0&1\\1&1\end{smallmatrix}\bigr) \oplus 0_\ell \oplus A$, so $\dim\Hull(C) = \ell$, and $\tau = \iota = 1$.
For independence, the other rows lie in $V_0$ and vanish in column $n$; a relation $a\mathbf{1} + b\, b_n \in \langle\text{other rows}\rangle$ forces $a\,\mathbf{1}_{\{1,\ldots,n-1\}} \in V_0$, hence $a = 0$ since $n - 1$ is odd, and then $b = 0$.

\emph{Case $(0,1)$, $n$ even.} The vector $\mathbf{1}$ is isotropic and orthogonal to every even-weight row, so the Gram matrix is $0_1 \oplus 0_{\ell-1} \oplus A$ and $\dim\Hull(C) = \ell$; all rows lie in $V_0$, so $\tau = 0$, and $\iota = 1$.
The rows are independent because only $\mathbf{1}$ is non-zero in column $n$.

In each case $\dim C = k$, and Lemma~\ref{lem:basic}(1) gives the hull dimension.
\end{proof}

\subsection{Uniform sampling of the pseudo-symplectic group}

Theorem~\ref{thm:witt} fails in $V$, so Lemma~\ref{lem:seq} cannot be applied in $V$ itself.
Instead, by Theorem~\ref{thm:ps-group} an isometry fixes $\mathbf{1}$, and for $n$ even it is determined by the image $u'$ of a vector $u$ with $\lambda(u) = 1$ together with a symplectic isomorphism $\langle\mathbf{1},u\rangle^\perp \to \langle\mathbf{1},u'\rangle^\perp$; for $n$ odd it is determined by its restriction to the symplectic space $V_0$.
The symplectic part is drawn by Lemma~\ref{lem:seq} exactly as in Proposition~\ref{prop:sp-exact}, with $U_i$ the orthogonal complement, inside $\langle\mathbf{1},u'\rangle^\perp$ (or $V_0$), of the vectors drawn so far.
The group sampler thus maps a fixed basis adapted to this decomposition onto a random one; we fix the following explicit basis.
Write $\mathbf{1}_J := \sum_{j \in J} b_j$ for $J \subseteq \{1, \ldots, n\}$.

\begin{lemma}[An explicit symplectic basis]\label{lem:ps-basis}
Let $\nu := \lfloor (n-1)/2 \rfloor$, and for $1 \le i \le \nu$ put $e_i := \mathbf{1}_{\{1, \ldots, 2i\}}$ and $f_i := b_{2i} + b_{2i+1}$.
Then $B(e_i,e_j) = B(f_i,f_j) = 0$ and $B(e_i,f_j) = \delta_{ij}$, so these vectors form a symplectic basis of a non-degenerate subspace $S \subseteq V_0$ of dimension $2\nu$, and $S^\perp = \langle \mathbf{1} \rangle$ for $n$ odd and $S^\perp = \langle \mathbf{1}, b_n \rangle$ for $n$ even.
\end{lemma}
\begin{proof}
All $e_i, f_i$ have even weight, so they lie in $V_0$.
Since $B(\mathbf{1}_J, \mathbf{1}_{J'}) = |J \cap J'| \bmod 2$, we get $B(e_i,e_j) = 2\min(i,j) = 0$, $B(f_i,f_j) = 0$ for $i \ne j$, and $B(e_i,f_j) = |\{1,\ldots,2i\} \cap \{2j,2j+1\}| \bmod 2 = \delta_{ij}$.
Hence $S$ is non-degenerate of dimension $2\nu$ and $\dim S^\perp = n - 2\nu$; $\mathbf{1} \perp S$ because $S \subseteq V_0$, and for $n$ even $b_n \perp S$ because the supports lie in $\{1, \ldots, n-1\}$.
\end{proof}

\begin{proposition}[Uniform pseudo-symplectic transformation]\label{prop:ps-exact}
Let $e_1, f_1, \ldots, e_\nu, f_\nu$ be the basis of Lemma~\ref{lem:ps-basis}.
For $n$ even, draw $x$ uniformly from $V$, put $u' := x + (1 - \lambda(x))\, b_n$, and draw a symplectic basis $e'_1, f'_1, \ldots, e'_\nu, f'_\nu$ of $\langle \mathbf{1}, u'\rangle^\perp$ by the recursion of Proposition~\ref{prop:sp-exact} started from $U_1 := \langle \mathbf{1}, u'\rangle^\perp$; for $n$ odd, draw such a basis of $V_0$, started from $U_1 := V_0$.
Then the map $g$ with $\mathbf{1} \mapsto \mathbf{1}$, $b_n \mapsto u'$ ($n$ even), $e_i \mapsto e'_i$, and $f_i \mapsto f'_i$ is exactly uniform on $\Ps(n,q)$, and it is obtained in an expected $O(n^3)$ operations in $\F_q$.
\end{proposition}
\begin{proof}
\emph{Uniformity.}
By Lemma~\ref{lem:ps-basis}, $\langle \mathbf{1}, b_n\rangle^\perp = S$ for $n$ even and $V_0 = S$ for $n$ odd, with symplectic basis $e_1, f_1, \ldots, e_\nu, f_\nu$.
Since $\lambda$ is linear and $\lambda(b_n) = 1$, $u'$ is uniform on the vectors with $\lambda(u') = 1$.
Given $u'$, Proposition~\ref{prop:sp-exact} gives a uniform symplectic basis of $\langle \mathbf{1}, u'\rangle^\perp$, so the map $e_i \mapsto e'_i$, $f_i \mapsto f'_i$ is uniform on the symplectic isomorphisms $S \to \langle \mathbf{1}, u'\rangle^\perp$.
By Theorem~\ref{thm:ps-group}(3) with $u := b_n$, $g$ is uniform on $\Ps(n,q)$; for $n$ odd, Theorem~\ref{thm:ps-group}(2) and Proposition~\ref{prop:sp-exact} give the same.

\emph{Cost.}
Drawing $u'$ costs $O(n)$, since it needs one uniform vector $x$ and the sum $\lambda(x)$; the symplectic basis costs an expected $O(n^3)$ by Proposition~\ref{prop:sp-exact}.
In matrices, $g = P^{-1} P'$, where the rows of $P$ are $\mathbf{1}, b_n, e_1, f_1, \ldots, e_\nu, f_\nu$ (without $b_n$ for $n$ odd) and those of $P'$ their images; $P^{-1}$ is computed once, and $P^{-1}P'$ costs $O(n^3)$.
\end{proof}

\begin{corollary}[Euclidean sampler, $q$ even]\label{cor:even-sampler}
For every admissible $(n,k,\ell)$, Algorithm~\ref{alg:main} with the orbits and their sizes from Theorem~\ref{thm:ps-orbits}, the representatives of Proposition~\ref{prop:ps-reps}, and the group sampler of Proposition~\ref{prop:ps-exact} returns an exactly uniform element of $\Cset^{\mathrm{E}}_{n,k,\ell}$ in an expected $O(k^2 n)$ operations in $\F_q$.
\end{corollary}
\begin{proof}
Uniformity follows from Theorem~\ref{thm:uniformity}, Theorem~\ref{thm:ps-orbits}, and Proposition~\ref{prop:ps-exact}.
The sampler draws the images of the basis $\mathbf{1}, b_n, e_1, f_1, e_2, f_2, \ldots$ in this order (without $b_n$ for $n$ odd).
In this basis $b_{2i-1} + b_{2i} = e_i + e_{i-1}$ (with $e_0 := 0$) and $b_{2i} + b_{2i+1} = f_i$, so every row $b_j + b_{j+1}$ of a block $H_a$ or $J_m$ with $j + 1 \le w$ is a combination of $e_1, f_1, \ldots, e_t, f_t$ with $t := \lceil (w-1)/2 \rceil$.
The representatives of Table~\ref{tab:ps-reps} consist of such rows, $\mathbf{1}$, and $b_n$, with block width $w \le k + \ell + 2$.
By Proposition~\ref{prop:trunc} the sampler can therefore be stopped after the first $t \le (k + \ell + 2)/2$ pairs, which cost an expected $O(k^2 n)$ operations as in the proof of Proposition~\ref{prop:sp-exact}.
\end{proof}

\section{Applications}\label{sec:applications}

\subsection{Code classification}

Classification of codes up to permutation equivalence usually proceeds by systematic enumeration, in which candidate generator matrices are visited in a fixed order and compared with a growing list of inequivalent representatives; equivalence classes that lie late in the order may never be reached in practice.
The samplers supply a complementary approach.
A permutation $\pi$ of $\{1, \ldots, n\}$ acts on $V$ by permuting coordinates, in the symplectic case simultaneously on both halves of $(a|b) \in \F_q^{2n}$; it preserves $B$, hence $\Cset_{n,k,\ell}$.
For a code $C$ let $\mathrm{PAut}(C)$ be the group of such permutations fixing $C$.

\begin{proposition}[Coverage]\label{prop:coverage}
Let $[C_1], \ldots, [C_M]$ be the permutation-equivalence classes in $\Cset_{n,k,\ell}$, and $p_j := n!/(|\mathrm{PAut}(C_j)|\,|\Cset_{n,k,\ell}|)$.
A uniform draw from $\Cset_{n,k,\ell}$ lies in $[C_j]$ with probability $p_j$, and $T \ge p_{\min}^{-1} \ln(M/\delta)$ independent draws, $p_{\min} := \min_j p_j$, meet every class with probability at least $1 - \delta$.
\end{proposition}
\begin{proof}
The class $[C_j]$ is the orbit of $C_j$ under these permutations, so it has $n!/|\mathrm{PAut}(C_j)|$ elements.
A fixed class is missed by all $T$ draws with probability $(1 - p_j)^T \le e^{-T p_{\min}} \le \delta/M$, and the union bound over the $M$ classes gives the claim.
\end{proof}

Completeness of a list of representatives is certified exactly by the mass formula: the list is complete if and only if $\sum_{[C]} n!/|\mathrm{PAut}(C)| = |\Cset_{n,k,\ell}|$, with the right-hand side given by the Anzahl theorems~\cite{Wan,Wan97}.
The two approaches complement each other: the classes hit least often are those with large automorphism groups, and once these highly symmetric codes have been classified by algebraic constructions, most of the remaining classes have trivial automorphism group and are hit with the largest probability $n!/|\Cset_{n,k,\ell}|$; with $p_{\min}$ taken over the remaining classes, Proposition~\ref{prop:coverage} then requires far fewer draws than for all classes.

\subsection{Public-key cryptography}

The Support Splitting Algorithm of Sendrier~\cite{Sendrier2000} solves the Permutation Equivalence Problem in time growing as $q^\ell$ in the hull dimension $\ell$, so code-based schemes whose security reduces to this problem prefer self-orthogonal codes, for which $\ell = k$ is largest.
The SPECK signature scheme~\cite{SPECK} samples a uniformly random self-orthogonal code for its key generation (\cite[Sec.~6, Def.~3]{SPECK}), and the Hollow~LWE scheme~\cite{HollowLWE} samples uniformly random codes with prescribed hull dimension.
SPECK points to the sampler of~\cite{HollowLWE}, which is proved uniform only for $2\ell \le k$.
The sampler of~\cite{ABL26} covers odd $q$; for $n$ and $q$ polynomial in $k$ it aborts with negligible probability and is otherwise exactly uniform for a prescribed hull dimension and square class~\cite[Thm.~28]{ABL26}, and for a prescribed hull dimension alone it is within total variation distance $q^{-(k-\ell)/2}$ of uniform~\cite[Cor.~30]{ABL26}, and weighting the square classes by their sizes makes it exactly uniform (Section~\ref{sec:intro}).
Corollaries~\ref{cor:euclid-sampler} and~\ref{cor:even-sampler} give exact uniformity in every characteristic; in particular, schemes over $\F_{2^m}$ with a prescribed hull dimension obtain a uniformity guarantee only from Corollary~\ref{cor:even-sampler}.
For odd $q$, our sampler costs an expected $O(k^2 n + k \log q)$ operations for all $n$, $k$, and $q$, against the bounds of~\cite{ABL26} recalled in Section~\ref{sec:intro}.

\subsection{Quantum error correction}\label{ssec:ea}

Under the standard correspondence between Pauli operators on $n$ qudits modulo phases and vectors of $\F_q^{2n}$, two operators commute if and only if their vectors are orthogonal for the trace-symplectic form~\cite{AK01}, \cite[Lemma~5]{KKKS06}; for $\F_q$-linear subspaces, the dual for this form coincides with the symplectic dual~\cite[proof of Thm.~2]{GHMR19}.
A subspace $C \le \F_q^{2n}$ of dimension $k$ and symplectic hull dimension $\ell$ is therefore the check space of an EA stabilizer code on $n$ qudits~\cite{BDH06,GHMR19} that consumes $c := (k-\ell)/2$ ebits and encodes $K := n - (k+\ell)/2$ qudits, and $c$ is the least number of ebits with which $C$ can be realized~\cite{WB08}, \cite[Thm.~2]{GHMR19}.
Through an $\F_q$-linear coordinatewise bijection $\F_{q^2}^n \to \F_q^{2n}$ that carries the trace-alternating form of~\cite[Sec.~IV]{KKKS06} to the trace-symplectic form~\cite[Lemma~14]{KKKS06}, an $\F_{q^2}$-linear code in $\Cset^{\mathrm{H}}_{n,k,\ell}$ likewise gives an EA stabilizer code that consumes $k - \ell$ ebits and encodes $n - k - \ell$ qudits~\cite{GHMR19}; for $\ell = k$ this is the construction of~\cite{CRSS98} for $q = 2$, extended to all $q$ in~\cite{AK01}, \cite[Cor.~19]{KKKS06}.
Through these correspondences, Corollaries~\ref{cor:sp-sampler} and~\ref{cor:u-sampler} are, without change, exactly uniform samplers of EA stabilizer codes with a prescribed number $K$ of encoded qudits and a prescribed number $c$ of ebits: take $k = n - K + c$ and $\ell = n - K - c$ in the symplectic case.
Previous exact samplers cover only $c = 0$, where a uniformly random Clifford operator applied to a fixed stabilizer code gives a uniform stabilizer code~\cite{KS14,BM21}.
Our samplers cover every $c$, and every $q$, in an expected $O(k^2 n + k \log q)$ operations (Corollaries~\ref{cor:sp-sampler} and~\ref{cor:u-sampler}), so EA codes with a prescribed number of ebits can be drawn in practice, for instance for Monte Carlo estimates of their properties, such as the minimum distance, the weight enumerator, or the automorphism group.

\section{Conclusion}\label{sec:conclusion}

We have given exactly uniform samplers for linear codes with prescribed hull dimension, for the Euclidean, Hermitian, and symplectic inner products and over finite fields of every characteristic.
All samplers follow one framework: the orbits of the isometry group with their sizes, a simple representative of each orbit, and an exactly uniform isometry.
The samplers cost an expected $O(k^2 N + k \log q)$ field operations per sample for codes of dimension $k$ in a space of dimension $N$, for all $n$, $k$, and $q$, and an implementation is available~\cite{hullsampler}.
Future directions include applying the samplers to the classification of codes, to the key generation of code-based schemes, and to the study of typical quantum error-correcting codes with a prescribed number of ebits.

\appendix
\section{Pseudo-symplectic geometry}\label{app:ps}

Theorem~\ref{thm:ps-group} is due to Dieudonn\'e~\cite{Dieudonne48} (see~\cite[p.~370]{Wan97}), and the order of $\Ps(n,q)$ is also in~\cite{MacWilliams69}.
Theorem~\ref{thm:ps-orbits} is due to Liu and Wan~\cite{LiuWan}; it is compiled in Wan's monograph~\cite{Wan}, which is not readily available, and apart from the survey~\cite{Wan97} we are not aware of other literature treating these results.
To keep the paper self-contained, we prove both theorems here; none of this is new.

We recall the notation.
Throughout, $q$ is even and $V := \F_q^n$ carries the Euclidean form $B(x,y) := \sum_i x_i y_i$, with standard basis $b_1, \ldots, b_n$.
Put $\lambda(x) := \sum_i x_i$ and $\mathbf{1} := (1, \ldots, 1)$, so that $B(\mathbf{1}, y) = \lambda(y)$ and, as $q$ is even, $B(x,x) = \lambda(x)^2$; let $V_0 := \ker\lambda = \mathbf{1}^\perp$.
$\Ps(n,q) := \{ g \in \GL(V) : B(xg,yg) = B(x,y) \ \forall x,y \}$ is the isometry group of $(V,B)$.
For a subspace $D$ of a space with a form $B$, $\Rad(D) := \{d \in D : B(d, D) = 0\}$ is the radical of $B|_D$; for a code $C \le V$ it is the hull $\Hull(C) = C \cap C^\perp$.
$\Cset_{n,k,\ell}$ is the set of codes $C \le V$ of dimension $k$ with $\dim\Hull(C) = \ell$, and for a code $C$ we put $\tau(C) := 0$ if $C \subseteq V_0$ and $\tau(C) := 1$ otherwise, and $\iota(C) := 1$ if $\mathbf{1} \in C$ and $\iota(C) := 0$ otherwise; $\Cset^{\tau,\iota}_{n,k,\ell}$ is the set of $C \in \Cset_{n,k,\ell}$ with this pair.
A space is symplectic if it carries a non-degenerate alternating form, $\Sp(U)$ is the isometry group of a symplectic space $U$, and $\Cset^{\mathrm{S}}_{\nu,k,\ell}$ is the set of $k$-dimensional subspaces with radical of dimension $\ell$ of a symplectic space of dimension $2\nu$ (its size does not depend on the space).
For a vector space $D$ we write $D^\vee := \mathrm{Hom}(D, \F_q)$.
Let $\nu := \lfloor (n-1)/2 \rfloor$, let $W := \langle \mathbf{1} \rangle$ for $n$ odd and $W := \langle \mathbf{1}, b_n \rangle$ for $n$ even, and let $S := W^\perp$.

\subsection{The group}

\begin{restated}{thm:ps-group}
\PsGroupStatement
\end{restated}

\begin{proof}
(1) For an isometry $g$, $\lambda(xg)^2 = B(xg,xg) = B(x,x) = \lambda(x)^2$, and squaring is injective in characteristic two, so $\lambda(xg) = \lambda(x)$ for all $x$; hence $V_0 g = V_0$, and $B(\mathbf{1}g, y) = B(\mathbf{1}g, (yg^{-1})g) = B(\mathbf{1}, yg^{-1}) = \lambda(yg^{-1}) = \lambda(y) = B(\mathbf{1},y)$ for all $y$, so $\mathbf{1}g = \mathbf{1}$ by non-degeneracy.

(2) $B(\mathbf{1},\mathbf{1}) = \lambda(\mathbf{1}) = n = 1$, so $\langle\mathbf{1}\rangle$ is non-degenerate, $V = \langle\mathbf{1}\rangle \perp V_0$ with $V_0 = \mathbf{1}^\perp$ non-degenerate of dimension $n-1$, and $V_0$ is alternating because $B(x,x) = \lambda(x)^2$.
Every isometry restricts to an element of $\Sp(V_0)$ by (1); conversely, every element of $\Sp(V_0)$ extends uniquely to an isometry of $V$ that fixes $\mathbf{1}$.

(3) Put $W_u := \langle\mathbf{1},u\rangle$.
Its Gram matrix is $\bigl(\begin{smallmatrix}0&1\\1&1\end{smallmatrix}\bigr)$, since $B(\mathbf{1},\mathbf{1}) = \lambda(\mathbf{1}) = 0$, $B(\mathbf{1},u) = \lambda(u) = 1$, and $B(u,u) = \lambda(u)^2 = 1$; so $W_u$ is non-degenerate, $V = W_u \perp W_u^\perp$, and $W_u^\perp$ is alternating of dimension $n - 2$, as every $x \in W_u^\perp$ has $\lambda(x) = B(\mathbf{1},x) = 0$.
An isometry $g$ with $ug = u'$ fixes $\mathbf{1}$ by (1), so it maps $W_u$ onto $W_{u'}$ and $W_u^\perp$ onto $W_{u'}^\perp$; hence $g$ is determined by $g|_{W_u^\perp}$, which is a symplectic isomorphism.
Conversely, $W_u$ and $W_{u'}$ have the same Gram matrix, so $\mathbf{1} \mapsto \mathbf{1}$, $u \mapsto u'$ together with any symplectic isomorphism $W_u^\perp \to W_{u'}^\perp$ is an isometry of $V = W_u \perp W_u^\perp$ onto $V = W_{u'} \perp W_{u'}^\perp$.
Symplectic isomorphisms $W_u^\perp \to W_{u'}^\perp$ exist by Theorem~\ref{thm:alt-classification}, so $\Ps(n,q)$ is transitive on the $q^{n-1}$ vectors $u'$ of $u + V_0$; for $u' = u$ the bijection identifies the stabilizer of $u$ with $\Sp(W_u^\perp) \cong \Sp(n-2,q)$, and the order follows.
\end{proof}

So $W$ is non-degenerate, $V = W \perp S$, and $S$ is a non-degenerate alternating space of dimension $2\nu$, by Theorem~\ref{thm:ps-group}(2) for $n$ odd (where $S = V_0$) and by Theorem~\ref{thm:ps-group}(3) with $u = b_n$ for $n$ even.
We write $\pi: V \to S$ for the projection along $W$.

\subsection{Two lemmas}

\begin{lemma}[Witt's theorem for $\Ps(n,q)$]\label{lem:app-witt}
Let $D, D' \le V$ and let $f: D \to D'$ be an isometry.
Then $f$ is the restriction of some $g \in \Ps(n,q)$ if and only if $f(\mathbf{1}) = \mathbf{1}$ when $\mathbf{1} \in D$, and $\mathbf{1} \notin D'$ when $\mathbf{1} \notin D$.
\end{lemma}
\begin{proof}
The conditions are necessary since every element of $\Ps(n,q)$ fixes $\mathbf{1}$.
Conversely, assume them.
The isometry $f$ preserves $\lambda$, as $\lambda(f(d))^2 = B(f(d),f(d)) = B(d,d) = \lambda(d)^2$, and squaring is injective as $q$ is even.
If $\mathbf{1} \notin D$, then $\mathbf{1} \notin D'$, and since $B(\mathbf{1},\cdot) = \lambda$, the map $f$ together with $\mathbf{1} \mapsto \mathbf{1}$ is an isometry $D \oplus \langle\mathbf{1}\rangle \to D' \oplus \langle\mathbf{1}\rangle$; so we may assume $\mathbf{1} \in D$.

If $n$ is odd, then $D = \langle \mathbf{1} \rangle \perp (D \cap S)$, and likewise for $D'$, so $f$ maps $D \cap S$ onto $D' \cap S$.
By Theorem~\ref{thm:witt}, $f|_{D \cap S}$ extends to an element of $\Sp(S)$, which extends to the required $g$ by Theorem~\ref{thm:ps-group}(2).

Let $n$ be even.
If $\lambda(D) = 0$, take any $u$ with $\lambda(u) = 1$; since $B$ is non-degenerate, some $u' \in V$ has $B(u', f(d)) = B(u, d)$ for all $d \in D$, and $d = \mathbf{1}$ gives $\lambda(u') = 1$.
As $\lambda(D) = \lambda(D') = 0$ and $B(u,u) = B(u',u') = 1$, setting $u \mapsto u'$ extends $f$ to an isometry $D \oplus \langle u \rangle \to D' \oplus \langle u' \rangle$, so we may assume $\lambda(D) \ne 0$.
Then some $u \in D$ has $\lambda(u) = 1$; put $u' := f(u)$, so $\lambda(u') = 1$.
With $W_u := \langle \mathbf{1}, u \rangle$, which is non-degenerate by Theorem~\ref{thm:ps-group}(3), $D = W_u \perp D_1$ with $D_1 := D \cap W_u^\perp$, and $f$ maps $W_u$ onto $W_{u'}$ and $D_1$ onto $D' \cap W_{u'}^\perp$.
As $W_u^\perp$ and $W_{u'}^\perp$ are symplectic of the same dimension, Theorems~\ref{thm:alt-classification} and~\ref{thm:witt} give a symplectic isomorphism $\sigma: W_u^\perp \to W_{u'}^\perp$ extending $f|_{D_1}$.
By Theorem~\ref{thm:ps-group}(3), $\mathbf{1} \mapsto \mathbf{1}$ and $u \mapsto u'$ together with $\sigma$ define some $g \in \Ps(n,q)$, and $g|_D = f$.
\end{proof}

\begin{lemma}[Rank-one changes]\label{lem:app-rank-one}
Let $D$ be a space with an alternating form $B$ and radical $R$.
\begin{enumerate}
\item For $\varphi \in D^\vee$ the form $B'(d,d') := B(d,d') + \varphi(d)\varphi(d')$ on $D$ has radical $R \cap \ker\varphi$, which is $R$ if $\varphi|_R = 0$ and a hyperplane of $R$ otherwise.
\item Let $F := D \oplus \langle c \rangle$ carry a symmetric bilinear form extending $B$ with $B(c,c) = 1$, and put $\psi := B(c,\cdot)|_D$.
Then $\Rad(F) = R \cap \ker\psi$.
\end{enumerate}
\end{lemma}
\begin{proof}
(1) Let $d \in \Rad(B')$.
Then $0 = B'(d,d) = B(d,d) + \varphi(d)^2 = \varphi(d)^2$, so $\varphi(d) = 0$ and $B(d,\cdot) = B'(d,\cdot) = 0$, that is, $d \in R \cap \ker\varphi$.
Conversely every $d \in R \cap \ker\varphi$ lies in $\Rad(B')$.
If $\varphi|_R \ne 0$, then $R \cap \ker\varphi$ is a hyperplane of $R$.

(2) Let $x = d + ac \in \Rad(F)$ with $d \in D$.
Then $0 = B(x,x) = B(d,d) + a^2 B(c,c) = a^2$ (as $q$ is even), so $x = d$, and $d \in \Rad(F)$ if and only if $B(d,\cdot)|_D = 0$ and $B(d,c) = \psi(d) = 0$, that is, $d \in R \cap \ker\psi$.
\end{proof}

\subsection{Proof of Theorem~\ref{thm:ps-orbits}}

For a code $C$ put $C_0 := C \cap V_0$; it is $C$ if $\tau(C) = 0$ and a hyperplane of $C$ if $\tau(C) = 1$, and $B|_{C_0}$ is alternating.
If $\tau(C) = 1$, every $c \in C$ with $\lambda(c) = 1$ satisfies $B(c,c) = 1$, and $C = C_0 \oplus \langle c \rangle$.

\begin{restated}{thm:ps-orbits}
\PsOrbitsStatement
\end{restated}

\begin{center}
\PsOrbitsTabular
\end{center}
\noindent Table~\ref{tab:ps-orbits} is reproduced above; for $n$ odd, $\Cset_{n,k,\ell}^{0,1}$ is empty.

\begin{proof}
Every element of $\Ps(n,q)$ preserves $\tau$ and $\iota$ by Theorem~\ref{thm:ps-group}(1), and $\tau(C) = 0$ if and only if $B|_C$ is alternating, as $B(x,x) = \lambda(x)^2$.

\emph{Single orbits.} Let $C, C' \in \Cset_{n,k,\ell}^{\tau,\iota}$.
By Lemma~\ref{lem:app-witt} it suffices to find an isometry $f: C \to C'$, with $f(\mathbf{1}) = \mathbf{1}$ when $\iota = 1$.
Write $C = \Rad(C) \perp N$ and $C' = \Rad(C') \perp N'$ with $N$ and $N'$ non-degenerate of dimension $k - \ell$ (Section~\ref{ssec:codes}).

Let $\iota = 0$; it suffices to find an isometry $N \to N'$, extended by any linear bijection $\Rad(C) \to \Rad(C')$.
If $\tau = 0$, then $N$ and $N'$ are alternating, and Theorem~\ref{thm:alt-classification} gives one.
If $\tau = 1$, then $N$ and $N'$ are not alternating, as $B(\rho + x, \rho + x) = B(x,x)$ for $\rho \in \Rad(C)$, and it suffices to show that a non-degenerate non-alternating space $N$ has an orthonormal basis; we use induction on $\dim N$.
Some $x \in N$ has $B(x,x) \ne 0$, and as squaring is a bijection of $\F_q$ we may assume $B(x,x) = 1$; then $N = \langle x \rangle \perp N_1$ with $N_1 := N \cap x^\perp$ non-degenerate.
If $N_1 = 0$ we are done, and if $N_1$ is not alternating, we apply induction to $N_1$.
Otherwise $N_1$ has a symplectic basis $e_1, f_1, \ldots, e_m, f_m$ by Theorem~\ref{thm:alt-classification}, and the vectors $x + e_1$, $x + f_1$, $x + e_1 + f_1$ are orthonormal, as $q$ is even, and span $\langle x, e_1, f_1 \rangle$; so $N = \langle x+e_1 \rangle \perp \langle x+f_1 \rangle \perp N_2$ with $N_2 := \langle x + e_1 + f_1, e_2, f_2, \ldots, e_m, f_m \rangle$ non-degenerate and not alternating, and we apply induction to $N_2$.

Let $\iota = 1$.
If $\tau = 0$ ($n$ is then even, as $\mathbf{1} \notin V_0$ for $n$ odd), then $\mathbf{1} \in C \subseteq \mathbf{1}^\perp$, so $\mathbf{1} \in \Rad(C)$, and likewise $\mathbf{1} \in \Rad(C')$; take for $f$ a linear bijection $\Rad(C) \to \Rad(C')$ fixing $\mathbf{1}$ together with an isometry $N \to N'$ (Theorem~\ref{thm:alt-classification}).
If $\tau = 1$ and $n$ is odd, then $B(\mathbf{1},\mathbf{1}) = 1$ and $C = \langle\mathbf{1}\rangle \perp C_0$ with $C_0$ alternating of dimension $k-1$ and $\Rad(C_0) = \Rad(C)$, and likewise for $C'$; take $\mathbf{1} \mapsto \mathbf{1}$ together with an isometry $C_0 \to C'_0$, which exists by Theorem~\ref{thm:alt-classification} applied to complements of the radicals.
If $\tau = 1$ and $n$ is even, pick $c \in C$ and $c' \in C'$ with $\lambda(c) = \lambda(c') = 1$; then $C = \langle\mathbf{1}, c\rangle \perp E$ with $\langle\mathbf{1},c\rangle$ non-degenerate by Theorem~\ref{thm:ps-group}(3) and $E := C \cap \langle\mathbf{1},c\rangle^\perp \subseteq V_0$ alternating of dimension $k-2$ with $\Rad(E) = \Rad(C)$, and likewise $C' = \langle\mathbf{1}, c'\rangle \perp E'$; take $\mathbf{1} \mapsto \mathbf{1}$, $c \mapsto c'$ together with an isometry $E \to E'$, which exists by Theorem~\ref{thm:alt-classification} applied to complements of the radicals.

\emph{Sizes.} We count each $\Cset_{n,k,\ell}^{\tau,\iota}$; the non-emptiness conditions are then read off from the sizes.

Suppose first that $n$ is odd.
Then $W = \langle\mathbf{1}\rangle$, $S = V_0$, and $\lambda(\mathbf{1}) = 1$, so every $x \in V$ is $x = \lambda(x)\mathbf{1} + \pi(x)$ with $\pi(x) \in S$, and $B(x,y) = \lambda(x)\lambda(y) + B(\pi(x),\pi(y))$.

\emph{Case $(0,1)$.} $\tau = 0$ means $C \subseteq S$, which does not contain $\mathbf{1}$; the set is empty.

\emph{Case $(0,0)$.} $\tau = 0$ means $C \subseteq S$, and then $\mathbf{1} \notin C$; so the set consists of the $k$-subspaces of $S$ with radical dimension $\ell$, and its size is $|\Cset^{\mathrm{S}}_{\nu,k,\ell}|$.

\emph{Case $(1,1)$.} $C = \langle \mathbf{1} \rangle \perp C_0$ with $C_0 = C \cap S$ of dimension $k-1$, and $\Rad(C) = \Rad(C_0)$ since $\mathbf{1}$ is anisotropic.
So $C \mapsto C_0$ is a bijection onto the $(k-1)$-subspaces of $S$ with radical dimension $\ell$, and the size is $|\Cset^{\mathrm{S}}_{\nu,k-1,\ell}|$.

\emph{Case $(1,0)$.} Here $\mathbf{1} \notin C$, so $\pi$ is injective on $C$, as its kernel is $\langle\mathbf{1}\rangle$, and $C = \{ d + \varphi(d)\mathbf{1} : d \in D \}$ with $D := \pi(C)$ and $\varphi := \lambda \circ (\pi|_C)^{-1} \in D^\vee$, where $\varphi \ne 0$ as $\tau = 1$; conversely each pair $(D,\varphi)$ with $\dim D = k$ and $\varphi \ne 0$ gives such a code.
In these coordinates $B|_C$ becomes $B(d,d') + \varphi(d)\varphi(d')$.
By Lemma~\ref{lem:app-rank-one}(1), with $R := \Rad(D)$ of dimension $r$ (so $k - r$ is even), the hull dimension is $\ell = r$ if $\varphi|_R = 0$ and $\ell = r - 1$ if $\varphi|_R \ne 0$.
The functionals with $\varphi|_R = 0$ correspond to $(D/R)^\vee$, so there are $q^{k-r}$ of them among the $q^k$ elements of $D^\vee$.
Hence for $k - \ell$ even the set consists of the pairs with $r = \ell$, $\varphi \ne 0$ and $\varphi|_R = 0$, that is, $q^{k-\ell} - 1$ functionals for each $D$, and for $k - \ell$ odd of the pairs with $r = \ell + 1$ and $\varphi|_R \ne 0$, that is, $q^k - q^{k-\ell-1}$ functionals for each $D$.
This gives the size in the table, in which the term of the other parity vanishes, as $\Cset^{\mathrm{S}}_{\nu,k',\ell'} = \emptyset$ when $k' - \ell'$ is odd.

Now let $n$ be even.
Then $V_0 = \langle\mathbf{1}\rangle \oplus S$, since $S \subseteq \mathbf{1}^\perp = V_0$ has codimension one in $V_0$ and $\mathbf{1} \notin S$, as $B(\mathbf{1},b_n) = 1$ while $B(S,b_n) = 0$.
So every $x \in V_0$ is $x = B(b_n,x)\mathbf{1} + \pi(x)$, and $B(x,y) = B(\pi(x),\pi(y))$ for $x, y \in V_0$, since $B(\mathbf{1}, V_0) = 0$.

\emph{Case $(0,1)$.} $C = \langle \mathbf{1} \rangle \oplus E$ with $E := C \cap S$ of dimension $k - 1$, and $\Rad(C) = \langle \mathbf{1} \rangle \oplus \Rad(E)$.
So $C \mapsto E$ is a bijection onto the $(k-1)$-subspaces of $S$ with radical dimension $\ell - 1$, and the size is $|\Cset^{\mathrm{S}}_{\nu,k-1,\ell-1}|$.

\emph{Case $(0,0)$.} $\pi$ is injective on $C$, $C = \{ d + \varphi(d)\mathbf{1} : d \in D\}$ with $D := \pi(C)$ of dimension $k$ and any $\varphi \in D^\vee$, and $\Rad(C) \cong \Rad(D)$; so there are $q^k |\Cset^{\mathrm{S}}_{\nu,k,\ell}|$ such codes.

\emph{Case $(1,1)$.} Here $C_0 \ni \mathbf{1}$ has dimension $k-1$, and $C_0 = \langle\mathbf{1}\rangle \oplus E_0$ with $E_0 := C_0 \cap S$ of dimension $k-2$, as $V_0 = \langle\mathbf{1}\rangle \oplus S$; conversely every $(k-2)$-subspace $E_0$ of $S$ gives such a $C_0 := \langle\mathbf{1}\rangle \oplus E_0$.
For each such $C_0$ there are $(q^{n-k+1}-1)/(q-1) - (q^{n-k}-1)/(q-1) = q^{n-k}$ subspaces $C \supseteq C_0$ of dimension $k$ not contained in $V_0$, and each of them has $C \cap V_0 = C_0$ and contains $\mathbf{1}$.
Pick $c \in C$ with $\lambda(c) = 1$ and put $\psi := B(c,\cdot)|_{C_0}$.
By Lemma~\ref{lem:app-rank-one}(2), $\Rad(C) = \Rad(C_0) \cap \ker\psi$; here $\Rad(C_0) = \langle\mathbf{1}\rangle \oplus \Rad(E_0)$, as $\mathbf{1} \in \Rad(C_0)$, and $\psi(\mathbf{1}) = \lambda(c) = 1$, so $\Rad(C)$ is a hyperplane of $\Rad(C_0)$ and $\ell = \dim\Rad(E_0)$.
So the orbit has $q^{n-k} |\Cset^{\mathrm{S}}_{\nu,k-2,\ell}|$ elements.

\emph{Case $(1,0)$.} Now $\mathbf{1} \notin C_0$, $\dim C_0 = k - 1$; let $R_0 := \Rad(C_0)$, of dimension $r$.
The codes $C \supseteq C_0$ with $\tau = 1$ are $C_0 \oplus \langle c \rangle$ with $c \in b_n + V_0$ determined modulo $C_0$, so there are $q^{n-k}$ of them, and none contains $\mathbf{1}$.
By Lemma~\ref{lem:app-rank-one}(2) with $\psi := B(c,\cdot)|_{C_0}$, $\ell = r$ if $\psi|_{R_0} = 0$ and $\ell = r - 1$ otherwise.
The map $V_0 \to R_0^\vee$, $v \mapsto B(v,\cdot)|_{R_0}$, is onto, since its kernel $(\langle \mathbf{1} \rangle + R_0)^\perp$ has dimension $n - 1 - r$; so as $c$ runs over $b_n + V_0$, $\psi|_{R_0}$ takes each value equally often, and $q^{n-k-r}$ of the $q^{n-k}$ codes have $\psi|_{R_0} = 0$.
By the case $(0,0)$ there are $q^{k-1}|\Cset^{\mathrm{S}}_{\nu,k-1,r}|$ choices of $C_0$, so
$|\Cset^{1,0}_{n,k,\ell}| = q^{n-k-\ell} q^{k-1} |\Cset^{\mathrm{S}}_{\nu,k-1,\ell}| + (q^{n-k} - q^{n-k-\ell-1}) q^{k-1} |\Cset^{\mathrm{S}}_{\nu,k-1,\ell+1}|$, as stated.

\emph{Non-emptiness.} Each size in the table is a sum of terms $a\,|\Cset^{\mathrm{S}}_{\nu,k',\ell'}|$ with $a > 0$, except that the coefficient $q^{k-\ell} - 1$ vanishes for $\ell = k$.
So $\Cset_{n,k,\ell}^{\tau,\iota}$ is non-empty if and only if some term with $a > 0$ has $\Cset^{\mathrm{S}}_{\nu,k',\ell'} \ne \emptyset$, that is, by Theorem~\ref{thm:sp-orbit}, $0 \le \ell' \le k'$, $k' - \ell'$ is even, and $k' + \ell' \le 2\nu$, where $2\nu = n-1$ for $n$ odd and $2\nu = n-2$ for $n$ even.
We use $k + \ell \equiv k - \ell \pmod 2$ throughout.
\begin{itemize}
\item $n$ odd, $(0,0)$: $(k',\ell') = (k,\ell)$ gives $k - \ell$ even and $k + \ell \le n-1$.
\item $n$ odd, $(1,1)$: $(k-1,\ell)$ gives $k - \ell$ odd and $k + \ell \le n$.
\item $n$ odd, $(1,0)$: $(k,\ell)$ with $\ell < k$ gives $\ell < k$, $k-\ell$ even, $k+\ell \le n-1$, and $(k,\ell+1)$ gives $k-\ell$ odd, $k+\ell \le n-2$; as $n-1$ is even, together $\ell < k$ and $k + \ell \le n-1$.
\item $n$ even, $(0,0)$: $(k,\ell)$ gives $k-\ell$ even and $k+\ell \le n-2$, that is, $k + \ell \le n-1$, as $k+\ell$ is even.
\item $n$ even, $(0,1)$: $(k-1,\ell-1)$ gives $\ell \ge 1$, $k-\ell$ even, and $k+\ell \le n$.
\item $n$ even, $(1,1)$: $(k-2,\ell)$ gives $k-\ell \ge 2$ even and $k+\ell \le n$.
\item $n$ even, $(1,0)$: $(k-1,\ell)$ gives $k-\ell$ odd and $k+\ell \le n-1$, and $(k-1,\ell+1)$ gives $k-\ell \ge 2$ even and $k+\ell \le n-2$; as $n-1$ is odd, together $\ell < k$ and $k+\ell \le n-1$.
\end{itemize}
These are the conditions in the table.
\end{proof}

\end{document}